\documentclass[11pt,letterpaper,reqno]{amsart}

\usepackage[T1]{fontenc}
\usepackage[utf8]{inputenc}
\usepackage{amsmath,amssymb,amsthm,mathtools}
\usepackage{bm}
\usepackage{enumitem}
\usepackage{booktabs,tabularx}
\usepackage{placeins}
\usepackage{needspace}
\usepackage[colorlinks=true,linkcolor=blue,citecolor=blue,urlcolor=blue]{hyperref}
\numberwithin{equation}{section}
\makeatletter
\@namedef{subjclassname@2020}{\textup{2020} Mathematics Subject Classification}
\makeatother
\theoremstyle{plain}
\newtheorem{theorem}{Theorem}[section]
\newtheorem{proposition}[theorem]{Proposition}
\newtheorem{lemma}[theorem]{Lemma}
\newtheorem{corollary}[theorem]{Corollary}
\theoremstyle{definition}

\theoremstyle{remark}
\newtheorem{remark}[theorem]{Remark}
\newcommand{\R}{\mathbb{R}}
\newcommand{\Z}{\mathbb{Z}}
\newcommand{\N}{\mathbb{N}}
\newcommand{\C}{\mathbb{C}}

\newcommand{\E}{\mathbb{E}}
\newcommand{\Prob}{\mathbb{P}}

\newcommand{\SL}{\mathrm{SL}}

\newcommand{\supp}{\operatorname{supp}}

\newcommand{\Var}{\operatorname{Var}}

\newcommand{\norm}[1]{\lVert #1 \rVert}
\newcommand{\abs}[1]{\lvert #1 \rvert}
\newcommand{\ip}[2]{\langle #1,#2\rangle}
\newcommand{\matr}[4]{
\begin{pmatrix}#1&#2\\#3&#4
\end{pmatrix}
}
\newcommand{\vect}[2]{
\begin{pmatrix}#1\\#2
\end{pmatrix}
}

\DeclareMathOperator{\spec}{spec}

\DeclareMathOperator{\ord}{ord}
\newcommand{\M}{\mathcal M}
\hypersetup{pdftitle={Exact transport exponents for random mosaic operators},pdfauthor={Fumihiko Nakano and Xu Xia}}
\begin{document}
\title[Transport for random mosaic operators]{Exact transport exponents for random mosaic operators}
\author{Fumihiko Nakano}
\address[Fumihiko Nakano]{Mathematical Institute, Tohoku University,\newline
6-3, Aramakiaza-Aoba, Aoba-ku, Sendai City, 980-8578, Japan}
\email{fumihiko.nakano.e4@tohoku.ac.jp}
\author[Xu Xia]{Xu Xia\textsuperscript{*}}
\address[Xu Xia]{State Key Laboratory of Mathematical Sciences,
Academy of Mathematics and Systems Science, Chinese Academy of Sciences,
100190 Beijing, P. R. China}
\email{xiaxu14@mails.ucas.ac.cn}
\thanks{\textsuperscript{*}Corresponding author: Xu Xia.}
\date{}
\begin{abstract}
We determine the exponentially time-averaged transport exponents of
period-two random mosaic operators, with zero potential on the even sites
and iid potential on the odd sites. For every bounded, strictly positive,
nonconstant single-site law, on one event of probability one,
\[
 \beta_f^-(q)=\beta_f^+(q)
 =\max\{0,2/3-\nu_H(f)/(3q)\}
\]
for all nonzero finitely supported states $f$ and all $q>0$.
The positive odd integer $\nu_H(f)$ is determined by divisibility by $H$
and cancellation against the bounded zero-energy solution. In particular,
the large-moment exponent is $2/3$ for every such state, while point sources
on the two sublattices have second-moment growth powers $1$ and $1/3$.
We obtain matching resolvent bounds by retaining the vanishing orders of
two polynomial source channels through the singular normalization at
zero energy. We also prove deterministic lower bounds for every bounded
positive configuration. No density assumption is imposed on the law.
\end{abstract}
\maketitle
\section{Introduction and main results}
\label{sec:intro}
Anderson's work \cite{Anderson} identified disorder as a mechanism for
the suppression of transport in a lattice. In the standard Anderson
model, the potential is random at every site. Rigorous localization
theory developed through, among other works, the one-dimensional
analysis of Kunz--Souillard \cite{KS} and the multiscale method of
Fr\"ohlich--Spencer \cite{FS} at large disorder or low energy.
Carmona--Klein--Martinelli \cite{CKM} established one-dimensional
localization for singular single-site distributions, including Bernoulli
laws. These results form a basic reference point for understanding how
the placement of randomness affects quantum motion.

Randomness restricted to a periodic sublattice leads to a different
problem. The intervening deterministic sites can produce exceptional
energies, and behavior near these energies may govern long-time
transport. We consider the simplest such geometry, with random and
zero potentials alternating along the lattice. Our question is to
determine the transport exponents, including their dependence on the
initial state.

\subsection{The random mosaic model}
Let $H=H_v$ be the bounded self-adjoint operator on $\ell^2(\Z)$ given by
\begin{equation}
\label{eq:mosaic-operator}
\begin{aligned}
 (H_vu)(n)&=u(n+1)+u(n-1)+V(n)u(n),\\
 V(2m)&=0,\qquad V(2m+1)=v_m.
\end{aligned}
\end{equation}
In the random setting we write $H_\omega$ and denote the common law of
the active-site potentials by $\mu$.
The transport statements use only bounded positivity,
\begin{equation}\label{rt:eq:law}
 v_m\text{ are iid},\qquad 0<a\le v_m\le b<\infty
 \quad\text{almost surely}.
\end{equation}
Here $\bar v=\E v_0>0$ and $\sigma_v^2=\Var(v_0)$; no density or
full-support hypothesis is needed. The lower bounds also allow a
constant law. We retain a separate deterministic
statement for $H=H_v$ whenever
\begin{equation}\label{tr:eq:assumption}
 0<a\leq v_m\leq b<\infty\qquad(m\in\Z).
\end{equation}
The latter bound does not require independence.

\subsection{Earlier work and the transport problem}
Periodic dilution of the Anderson model predates the terminology
``mosaic.'' Hilke \cite{Hilke} studied the localization properties of
a chain with random potentials on periodically spaced sites, identifying
the special role of resonant energies. De Moura and collaborators
\cite{deMoura} subsequently investigated eigenstates and wave-packet
spreading in diluted Anderson chains. More recently, Wang and
collaborators \cite{WangMosaic} established exact mobility edges for
\emph{quasiperiodic} mosaic lattices. That setting shares the periodic
placement of the active sites, but not the independent randomness in
\eqref{rt:eq:law}.

For disordered mosaic lattices, Nguyen--Kim \cite{NguyenKim} studied
transmission, participation ratios, and wave-packet dynamics numerically.
Their dynamical calculations use centered uniform disorder and Gaussian
initial packets; these differ from the positive laws and finitely
supported states considered here. Longhi \cite{LonghiMosaic} studied
resonances and gap-protected localization for generalized mosaic
lattices with a periodic background. The transport problem considered
here asks for time-averaged moment exponents of the physical quantum
evolution, rather than a criterion for resonant energies or mobility edges.

Rigorous anomalous transport is also known in random polymer models.
For their elliptic critical energies, Jitomirskaya--Schulz-Baldes--Stolz
\cite{JSS} obtained transport lower bounds. Under the nonresonance and
nondegeneracy assumptions of \cite[Theorem~3]{JSUpper}, the matching
upper bound gives the disorder-averaged point-source law
$\beta(q)=\max\{0,1-1/(2q)\}$. In our model the critical cell matrices
at zero energy are instead common triangular shears. The random
Kronig--Penney model provides a precedent for transport lower bounds
at such a parabolic critical energy \cite{DKSB}.

Here we obtain matching bounds for every finitely supported initial
state, on a common event of probability one. The propagation scale
is the same for all these states, but their coupling to the threshold
is not: even point sources at adjacent sites have different moment
exponents. The polynomial representation below records this dependence
without imposing regularity on the single-site law.

\subsection{Transport exponents and the main result}
For $q>0$, $T>0$, and finitely supported $f\ne0$, define
\begin{equation}\label{tr:eq:moment}
 \M_q(T;f)=\frac2T\int_0^\infty e^{-2t/T}
 \sum_{n\in\Z}|n|^q\abs{\ip{\delta_n}{e^{-itH}f}}^2\,dt.
\end{equation}

Set
\begin{equation}\label{rt:eq:beta}
 \beta_f^-(q)=\liminf_{T\to\infty}
       \frac{\log\M_q(T;f)}{q\log T},\qquad
 \beta_f^+(q)=\limsup_{T\to\infty}
       \frac{\log\M_q(T;f)}{q\log T},
\end{equation}
and $\alpha_u^\pm(f)=\lim_{q\to\infty}\beta_f^\pm(q)$.
H\"older's inequality makes $\beta_f^\pm$ nondecreasing; the ballistic
bound makes these large-moment limits finite.
There are unique polynomials $A_f,B_f\in\C[E]$ such that
\begin{equation}\label{tr:eq:module}
 f=A_f(H)\delta_0+B_f(H)\delta_1;
\end{equation}
see Lemma~\ref{tr:lem:module}. For a nonzero polynomial $P$, let
$\ord_0P$ be the smallest power of $E$ with nonzero coefficient in $P(E)$;
set $\ord_0 0=+\infty$. Define
\begin{equation}\label{tr:eq:index}
 r_f=\ord_0 A_f,\qquad s_f=\ord_0 B_f,\qquad
 \nu_H(f)=\min\{4r_f+1,\,4s_f+3\}.
\end{equation}
This finite positive odd integer is the \emph{threshold index} of $f$.
Proposition~\ref{tr:prop:intrinsic} characterizes it without choosing
polynomial anchors.
For a fixed coordinate vector $f$, the polynomials in
\eqref{tr:eq:module}, and hence $\nu_{H_\omega}(f)$, may depend on the
sample. All assertions involving this index are pathwise; the
point-source indices below are deterministic.

\begin{theorem}\label{up:thm:front}
Assume \eqref{rt:eq:law} and $\sigma_v^2>0$. On one event of probability
one, for every nonzero finitely supported $f$ and every $q>0$,
\begin{equation}\label{up:eq:front}
 \beta_f^-(q)=\beta_f^+(q)
 =\max\left\{0,\frac23-\frac{\nu_H(f)}{3q}\right\}.
\end{equation}
In particular, $\alpha_u^-(f)=\alpha_u^+(f)=2/3$.
\end{theorem}
For point sources, the index is $1$ on the even sublattice and $3$ on
the odd sublattice; see Corollary~\ref{tr:cor:sites}. Thus, on the same
event, for every $j\in\Z$,
\begin{equation}\label{up:eq:second-exponents}
 \lim_{T\to\infty}\frac{\log\M_2(T;\delta_{2j})}{\log T}=1,
 \qquad
 \lim_{T\to\infty}\frac{\log\M_2(T;\delta_{2j+1})}{\log T}=\frac13.
\end{equation}
These are growth powers of the second moment: in the normalization
\eqref{rt:eq:beta}, the corresponding exponents are $1/2$ and $1/6$.
The transport exponents for the states $\delta_j$ do not depend on the
single-site law within the stated class; the constants need not be uniform
as the variance tends to zero.

More generally, the index has an intrinsic interpretation. Divide $f$ by the largest
power of $H$ that leaves a finitely supported state, then pair the
remaining state with the bounded zero-energy solution. The divisibility
order and the vanishing or nonvanishing of this pairing determine
$\nu_H(f)$; see Proposition~\ref{tr:prop:intrinsic}. Higher cancellations
occur, for example, in $H^k\delta_0$ and $H^k\delta_1$, whose indices
are $4k+1$ and $4k+3$.

The theorem identifies moment exponents, not leading asymptotic
constants or a pointwise-in-time diffusion law. In particular, an exponent of
zero does not imply bounded moments. Positive variance is essential:
Remark~\ref{up:rem:periodic} gives the periodic counterexample. The
quantitative lower bounds below do not require this nondegeneracy.

\subsection{Quantitative lower bounds and the two sublattices}
The next theorem supplies the lower half of
Theorem~\ref{up:thm:front}, with an explicit almost-sure loss, and a
separate estimate for disorder averages. Unlike the exact formula, it
also allows a constant single-site law.
\begin{theorem}\label{rt:thm:main}
Assume \eqref{rt:eq:law}.
\begin{enumerate}[label=\textup{(\roman*)},leftmargin=*]
\item There is an event $\Omega_*$ of probability one such that, for every
$\omega\in\Omega_*$, every nonzero finitely supported $f$, and every $q>0$,
there exist $c>0$ and $T_0>e^e$ with
\begin{equation}\label{rt:eq:quenched}
 \M_q(T;f)\ge cT^{(2q-\nu_H(f))/3}
                 (\log\log T)^{-(q+\nu_H(f))/3}
 \qquad(T\ge T_0).
\end{equation}
The constants may depend on the sample, $f,q$, and the law.
\item For each fixed $j\in\Z$ and $q>0$, deterministic constants $c>0$ and
$T_0<\infty$ satisfy
\begin{equation}\label{rt:eq:annealed}
 \begin{aligned}
 \E\M_q(T;\delta_{2j})&\ge cT^{(2q-1)/3},\\
 \E\M_q(T;\delta_{2j+1})&\ge cT^{(2q-3)/3}
 \end{aligned}
 \qquad(T\ge T_0).
\end{equation}
\end{enumerate}
\end{theorem}

The threshold indices are $\nu_H(\delta_0)=1$ and $\nu_H(\delta_1)=3$; their
algebraic verification is included in the proof of
Corollary~\ref{tr:cor:sites}. In particular, on $\Omega_*$, for each
fixed $j$,
\begin{equation}\label{rt:eq:second}
 \M_2(T;\delta_{2j})\ge\frac{cT}{\log\log T},
 \qquad
 \M_2(T;\delta_{2j+1})\ge
       \frac{cT^{1/3}}{(\log\log T)^{5/3}}
\end{equation}
for all sufficiently large $T$. Under positive variance these lower
bounds have the growth powers in \eqref{up:eq:second-exponents}; the
double-logarithmic factors are not claimed to be optimal.

\begin{theorem}\label{tr:thm:main}
Assume \eqref{tr:eq:assumption}. For each nonzero finitely supported $f$ and
each $q>0$, there are $c>0$ and $T_0<\infty$ such that
\begin{equation}\label{tr:eq:transport}
 \M_q(T;f)\geq cT^{(q-\nu_H(f))/2}\qquad(T\geq T_0).
\end{equation}
The constants may depend on $a,b,f,q$ and the coefficients of $A_f,B_f$.
Consequently, with \eqref{rt:eq:beta},
one has $\beta_f^-(q)\geq\frac12-\frac{\nu_H(f)}{2q}$.
\end{theorem}

The deterministic bound is weaker but holds for every positive bounded
configuration. Under iid randomness the gain in the power is
$(q+\nu_H(f))/6$, with an almost-sure double-logarithmic loss.
The deterministic sufficient condition for growing moments is
$q>\nu_H(f)$. Under nondegenerate iid randomness, positive power-law
growth occurs exactly when $2q>\nu_H(f)$, by
Theorem~\ref{up:thm:front}; subpower growth below this threshold is not
classified.

\subsection{The propagation scale and the source channels}
The power $2/3$ comes from two restrictions on propagation near zero
energy. Write $E=-\kappa^2$ and let $N$ count two-site cells. After a
singular rescaling, the centered random part of one transfer matrix
has size $O(\kappa)$; its accumulated quadratic variation has size
$O(N\kappa^2)$. Passing to the resolvent energy $E+i/T$ introduces
a further error of size $O((T\kappa)^{-1})$ per cell. On prefixes with
bounded normalized transfers and inverses, complex-energy stability
therefore requires $N/(T\kappa)$ to be small. Balancing these restrictions
gives
\begin{equation}\label{up:eq:scale-balance}
 N\asymp T\kappa,\qquad N\kappa^2\asymp1
 \quad\Longrightarrow\quad
 \kappa\asymp T^{-1/3},\qquad N\asymp T^{2/3}.
\end{equation}
The moment exponent also depends on the physical resolvent mass at this
distance. The singular normalization weights the two source channels
differently. We retain both polynomial factors in
\eqref{tr:eq:tail-module}, including their zeros at the threshold.
The two-sided observation estimate \eqref{rt:eq:shell-bridge} converts
the full source jump into mass on distant shells. It does not require
independence of a random boundary value and the transfer products.

For the upper bound, a phase corrector gives positive mean radial drift
uniformly in the starting angle. We combine the resulting product
growth with the full-line criterion of Damanik--Tcheremchantsev
\cite{DTUpper}; Lemma~\ref{up:lem:DT} specifies the conversion from
physical sites to cells. A weighted Schur estimate covers the shrinking
energy window where this growth is insufficient. Finally, we extract
the polynomial factors before Poisson smoothing and use a spatial
cutoff that depends on energy. This order of operations preserves the
threshold index in the upper bound.

The Schur complement has divergence form. Related quantum-dynamical
bounds for fixed divergence-type random Jacobi operators were obtained
by Li--Wang--Zhang \cite[Theorems~1.1--1.2]{LWZ}: their upper and lower
bounds in expectation do not match, and they also prove an almost-sure
lower bound. Here the Schur coefficients depend on the spectral
parameter. The reduction is therefore not a fixed unitary equivalence
of the quantum evolutions, and we estimate the physical mosaic resolvent.

Section~\ref{sec:prelim} establishes the polynomial representation,
the intrinsic threshold index, and the Schur identities used in both
bounds. Section~\ref{sec:transport} proves the deterministic lower
bound. Section~\ref{sec:random-transport} uses random cancellation to
obtain Theorem~\ref{rt:thm:main}. The transfer-growth estimates are
proved in Section~\ref{sec:transfer-growth}; Section~\ref{sec:upper-transport}
uses them to obtain the upper bound and complete the proof of
Theorem~\ref{up:thm:front}.

\paragraph{Relation to the companion paper.}
The companion manuscript \cite{NXStatistics} treats finite-volume
eigenvalue statistics: Poisson limits, fixed-edge half-clocks with
Gaussian corrections, moving clocks, and critical diffusion limits,
under the hypotheses stated there. The papers grew out of a common
draft and share the model, elementary transfer normalization, and
local phase calculations. The present transport proofs are
self-contained and use no statistical limit theorem.
The polynomial-channel bookkeeping is also related to the unpublished
manuscript \emph{Pure Point Spectrum and Anomalous Transport in Mosaic
Stark Operators}; no result from that manuscript is a proof input.

\section{Preliminaries and threshold channels}
\label{sec:prelim}
Throughout this section, $H=H_v$ satisfies \eqref{tr:eq:assumption},
unless randomness is explicitly required.
Vector norms are Euclidean and matrix norms are induced operator norms
unless a Frobenius norm is specified. Inner products are conjugate-linear
in the first argument. Transfer lengths count two-site cells; position
moments use the original lattice sites. Constants may change between
estimates, with dependence and uniformity specified where needed.
\subsection{Transfer conventions}
For $Hu=Eu$, write
\begin{equation}\label{eq:M-intro}
 T(E,x)=\matr{E-x}{-1}{1}{0},\qquad M(E,x)=T(E,0)T(E,x).
\end{equation}
Our convention is
\begin{equation}\label{eq:transfer-convention}
 \vect{u(n+1)}{u(n)}
 =T(E,V(n))\vect{u(n)}{u(n-1)}.
\end{equation}
Thus $M(E,v_m)$ propagates from $(u(2m+1),u(2m))^t$ to
$(u(2m+3),u(2m+2))^t$. At zero energy,
$-M(0,x)=I+xE_{21}$, where $E_{ij}$ are the matrix units.
This matrix maps $(s,t)^t$ to $(s,t+xs)^t$ and fixes $\R(0,1)^t$.
For $x>0$ it is not diagonalizable, and all critical matrices have
the same fixed line. We also write
$J=E_{21}-E_{12}=\matr{0}{-1}{1}{0}$.

\subsection{Polynomial channels and time averaging}
\begin{lemma}\label{tr:lem:module}
Every finitely supported $f$ has a unique representation
\eqref{tr:eq:module}. For $z\notin\spec H$, write
\[
 R(z)=(H-z)^{-1},\qquad G(n,j;z)=\ip{\delta_n}{R(z)\delta_j}.
\]
There is a finite set $F_f\subset\Z$, independent of $z$, such that
\begin{equation}\label{tr:eq:tail-module}
 (R(z)f)(n)=A_f(z)G(n,0;z)+B_f(z)G(n,1;z)
 \quad(n\notin F_f).
\end{equation}
\end{lemma}

\begin{proof}
The recurrence
$\delta_{n+1}=(H-V(n))\delta_n-\delta_{n-1}$, used in both directions,
generates every $\delta_n$ from $\delta_0,\delta_1$ over $\C[H]$.
For uniqueness, a polynomial $A$ of degree $d$ with a nonzero leading
coefficient gives $A(H)\delta_0$ nonzero at both extreme sites $-d,d$;
similarly $B(H)\delta_1$ of degree $e$ has extremes $1-e,1+e$.
If both are nonzero and $d\geq e+1$, the site $-d$ cannot cancel.
If $e\geq d$, the site $1+e$ cannot cancel. If only one polynomial is
nonzero, either of its extreme coefficients already proves that its
image is nonzero.

For any polynomial $P$, polynomial division gives
$P(H)-P(z)=(H-z)Q_z(H)$, where $\deg Q_z\leq\deg P-1$.
Thus $R(z)P(H)\delta_j=P(z)R(z)\delta_j+Q_z(H)\delta_j$.
The second term is supported in a fixed finite interval, proving
\eqref{tr:eq:tail-module}.
\end{proof}

\begin{proposition}\label{tr:prop:intrinsic}
Let $c_{00}(\Z)$ denote the finitely supported sequences, and let
$f\in c_{00}(\Z)\setminus\{0\}$. There is a largest integer $k\ge0$
such that $f\in H^k c_{00}(\Z)$. The finite sequence $g$ satisfying
$f=H^kg$ is unique, and
\begin{equation}\label{tr:eq:intrinsic}
 \nu_H(f)=
 \begin{cases}
 4k+1,&\displaystyle\sum_{m\in\Z}(-1)^m g(2m)\ne0,\\[2pt]
 4k+3,&\displaystyle\sum_{m\in\Z}(-1)^m g(2m)=0.
 \end{cases}
\end{equation}
For any $\ell\in\Z$, write instead
$f=A_f^{(\ell)}(H)\delta_{2\ell}+B_f^{(\ell)}(H)\delta_{2\ell+1}$.
Then
\[
 \min\{4\ord_0A_f^{(\ell)}+1,4\ord_0B_f^{(\ell)}+3\}=\nu_H(f).
\]
\end{proposition}
\begin{proof}
By Lemma~\ref{tr:lem:module}, $f\in H^j c_{00}(\Z)$ precisely when
$E^j$ divides both $A_f$ and $B_f$. Thus $k=\min(r_f,s_f)<\infty$. Define
\[
 \widehat A(E)=E^{-k}A_f(E),\quad \widehat B(E)=E^{-k}B_f(E),\qquad
 g=\widehat A(H)\delta_0+\widehat B(H)\delta_1.
\]
These are polynomials and $H^kg=f$. The same uniqueness shows that
$H^k$ is injective on $c_{00}(\Z)$, so $g$ is unique. Its polynomials
are $A_g=\widehat A$, $B_g=\widehat B$ and cannot both vanish at zero.

The real sequence $u_0(2m)=(-1)^m$, $u_0(2m+1)=0$ solves $Hu_0=0$.
Pairing it with a finite sequence uses only a finite sum, and symmetry
of the recurrence gives
$\sum_n u_0(n)(Hp)(n)=0$ for every $p\in c_{00}(\Z)$.
Therefore $\sum_nu_0(n)g(n)=A_g(0)$.
If this number is nonzero, $\ord_0 A_g=0$ and $\nu_H(g)=1$.
If it is zero, then $B_g(0)\ne0$ and $\nu_H(g)=3$.
Multiplication by $H^k$ raises both polynomial orders by $k$ and the
threshold index by $4k$, proving the formula. The integer $k$ and the
state $g$ do not involve the choice of the two generating sites.
Normalizing the zero-energy solution at $2\ell$ gives
$u_0^{(\ell)}=(-1)^\ell u_0$, and hence
\[
 A_g^{(\ell)}(0)=\sum_m(-1)^{m-\ell}g(2m)
               =(-1)^\ell A_g^{(0)}(0).
\]
Whether this number vanishes is therefore independent of $\ell$.
\end{proof}

\begin{lemma}\label{tr:lem:parseval}
For $\varepsilon=T^{-1}$,
\begin{equation}\label{tr:eq:parseval}
 \M_q(T;f)=\frac1{\pi T}\int_\R
 \sum_{n\in\Z}|n|^q\abs{(R(E+i\varepsilon)f)(n)}^2\,dE.
\end{equation}
All moments in \eqref{tr:eq:moment} are finite for finite $T$.
\end{lemma}

\begin{proof}
The Fourier transform, with kernel $e^{iEt}$, of
$\mathbf1_{t\geq0}e^{-t/T}\ip{\delta_n}{e^{-itH}f}$ is
$-i\ip{\delta_n}{R(E+i/T)f}$.
Scalar Parseval gives the factor $1/(\pi T)$ in
\eqref{tr:eq:parseval}. Summation follows by monotone convergence.

Write $H=V+K$, where $K$ is the hopping operator, and set
$U_I(t)=e^{itV}e^{-itH}$ and $K_I(t)=e^{itV}Ke^{-itV}$.
The equation $U_I'=-iK_I(t)U_I$, $U_I(0)=I$, has a norm-convergent
iterated integral expansion since $\|K_I(t)\|=2$.
A term with $k$ hops has at most $2^k$ paths and integration volume
$|t|^k/k!$. Each hopping entry has modulus one, and reaching $n$ from
$j$ requires $k\ge |n-j|$. Thus, for every $\alpha>0$,
\[
 \abs{\ip{\delta_n}{e^{-itH}\delta_j}}
 \leq\sum_{k\ge|n-j|}\frac{(2|t|)^k}{k!}
 \leq \exp\bigl(2e^\alpha |t|-\alpha|n-j|\bigr).
\]
Finite support gives the corresponding bound for $f$.
Split the position sum at $|n|=C_f+C_\alpha|t|$, with
$C_\alpha>2e^\alpha/\alpha$. Unitarity bounds the inner part by
$C(1+|t|)^q$, and the displayed estimate bounds the outer part by the same
quantity. Exponential time averaging is therefore finite.
\end{proof}

\subsection{A deterministic resolvent identity on the gap side}
\label{sec:gap-schur}
For later use in quantum dynamics, take an arbitrary configuration satisfying
\eqref{tr:eq:assumption} and write $H=H_v$. The following block identities
hold without randomness and give the gap-side resolvent input used below.
Define the unitary map
$\mathcal T:\ell^2(\Z)\to\ell^2(\Z)\oplus\ell^2(\Z)$ by
\[
 u(2m)=(-1)^m\psi_m,\qquad u(2m+1)=(-1)^m\chi_m.
\]
Let $D\psi(m)=\psi_m-\psi_{m+1}$ and
$D^*\chi(m)=\chi_m-\chi_{m-1}$. In these coordinates,
\begin{equation}\label{tr:eq:block}
 \widetilde H=\mathcal T H\mathcal T^{-1}
 =\begin{pmatrix}0&D^*\\D&V\end{pmatrix},\qquad
 V=\operatorname{diag}(v_m).
\end{equation}
Write $L_0=D^*D=DD^*$ for the positive discrete Laplacian.
For $0<E<a$, define
\begin{equation}\label{tr:eq:schur}
 c_m(E)=(v_m-E)^{-1},\quad
 Q_E=E+D^*c(E)D,\quad A_E=L_0+E(V-E).
\end{equation}
Here $Q_E\ge EI$ and $A_E\ge E(a-E)I$. To solve
$(\widetilde H-E)(\psi,\chi)=(F,G)$, first substitute
$\chi=(V-E)^{-1}(G-D\psi)$ to obtain
$-Q_E\psi=F-D^*(V-E)^{-1}G$. Alternatively, substitution of
$\psi=(D^*\chi-F)/E$ gives $A_E\chi=EG+DF$.
Solving these equations and substituting back gives
\begin{equation}\label{tr:eq:blocks}
 \begin{split}
 ((\widetilde H-E)^{-1})_{\psi\psi}&=-Q_E^{-1},\\
 ((\widetilde H-E)^{-1})_{\chi\chi}&=E A_E^{-1},\\
 ((\widetilde H-E)^{-1})_{\psi\chi}&=D^*A_E^{-1}.
 \end{split}
\end{equation}
In particular $(0,a)\cap\spec H=\varnothing$.

\begin{proposition}\label{tr:prop:spectrum}
Let $A=\supp\mu\subset(0,\infty)$ be compact. Under \eqref{rt:eq:law},
almost surely
\begin{equation}\label{tr:eq:spectrum}
 \spec H_\omega=\bigcup_{c\in A}
 \bigl([E_-(c),0]\cup[c,E_+(c)]\bigr),\qquad
 E_\pm(c)=\frac{c\pm\sqrt{c^2+16}}2.
\end{equation}
If $A=[a,b]$, this becomes
$[E_-(a),0]\cup[a,E_+(b)]$. For disconnected $A$ the union on the
positive side need not be an interval.
\end{proposition}
For finite $A$, this is a specialization of
\cite[Theorem~3.3.1 and Remark~3.3.2]{DFG}. We include the argument for
compact $A$ so that no interval-support assumption is imposed.
\begin{proof}
For $E\ne0$, eliminating the even-site components identifies
invertibility of $H-E$ with that of $K_E-(E^2-2)$, where
$(K_E\phi)_m=\phi_{m+1}+\phi_{m-1}+Ev_m\phi_m$.
The constant-$c$ operator has discriminant $E(E-c)-2$ and spectrum
$\Sigma_c=[E_-(c),0]\cup[c,E_+(c)]$. If $E$ lies outside the right
side of \eqref{tr:eq:spectrum}, then
$\inf_{c\in A}|E^2-2-Ec|>2$. The diagonal part of
$K_E-(E^2-2)$ has inverse norm less than $1/2$, whereas the hopping term
has norm two. A Neumann series proves invertibility for every
configuration with values in $A$.

For the converse, fix a countable dense subset $A_0\subset A$.
For each $c\in A_0$, length $l\in\N$ and accuracy $1/k$, an iid
configuration contains infinitely many disjoint $l$-cell blocks with
$|v_m-c|<1/k$, almost surely: each such block has the same positive
probability, and the disjoint-block events are independent. Intersect
these events over the countable choices. For $E\in\Sigma_c$, truncate
a bounded Bloch solution for the constant-$c$ operator to one of these
blocks. Its norm is comparable to $\sqrt l$; the two boundary errors
are bounded, and the potential error after normalization is at most
$1/k$. Taking $l,k\to\infty$ gives a Weyl sequence. The same argument
applies to the bounded band-edge solutions, including $E=0$.
Thus $\Sigma_c\subset\spec H_\omega$ for every $c\in A_0$.
Continuity of the interval endpoints in $c$ and closedness of the
spectrum give the inclusion for all $c\in A$.
\end{proof}

\section{Deterministic transport lower bounds}
\label{sec:transport}
We prove Theorem~\ref{tr:thm:main} and Corollary~\ref{tr:cor:sites} for
every deterministic configuration satisfying \eqref{tr:eq:assumption}.
For real $0<E<a$, the two columns $(H-E)^{-1}\delta_0$ and
$(H-E)^{-1}\delta_1$ have different sizes near zero. We prove the
real-energy bounds for these columns and extend them to a thin
complex-energy cone. The polynomial identity
\eqref{tr:eq:tail-module} then gives the bound for an arbitrary finite state.

Constants in the Green-function estimates below depend only on $a,b$,
unless otherwise indicated. All small-energy statements hold on a common
interval $0<E<E_0(a,b)$.
\subsection{The zero-site channel}

\begin{lemma}\label{tr:lem:conductance}
Suppose $c_-\leq c_m\leq c_+$ with $c_->0$, and let
$Q=E+D^*cD$, $0<E\leq1$. There are constants $c,C>0$, depending only on
$c_-,c_+$, such that
\begin{equation}\label{tr:eq:q-bounds}
 cE^{-1/2}e^{-C\sqrt E|m-j|}
 \leq (Q^{-1})_{mj}\leq CE^{-1/2}\qquad(m,j\in\Z).
\end{equation}
\end{lemma}

\begin{proof}
The inverse $Q^{-1}$ preserves coordinatewise nonnegativity.
Choose $\Lambda>\norm Q$. Since $Q\ge EI$ and its off-diagonal
entries are nonpositive, $I-Q/\Lambda$ has nonnegative entries and
norm at most $1-E/\Lambda<1$. Thus
\[
 Q^{-1}=\Lambda^{-1}\sum_{k\ge0}(I-Q/\Lambda)^k
\]
converges in operator norm and has nonnegative entries.

Inversion reverses quadratic-form order: if $A\ge B\ge cI>0$, then
$C=B^{-1/2}AB^{-1/2}\ge I$, and
$A^{-1}=B^{-1/2}C^{-1}B^{-1/2}\le B^{-1}$.
Apply this to $Q\geq E+c_-L_0$ and test on $\delta_j$ to obtain
\[
 (Q^{-1})_{jj}\leq((E+c_-L_0)^{-1})_{00}
 =\frac1{\sqrt{E(E+4c_-)}}\leq CE^{-1/2}.
\]
The equality follows by integrating
$(E+2c_--2c_-\cos\theta)^{-1}$ over $[-\pi,\pi]$.
Cauchy--Schwarz for the positive operator $Q^{-1}$ gives the upper
off-diagonal bound.

For the lower bound, it suffices to take $j=0$. Define the resistance
distance $s_0=0$ and
\[
 s_m=\sum_{k=0}^{m-1}c_k^{-1}\ (m>0),\qquad
 s_m=\sum_{k=m}^{-1}c_k^{-1}\ (m<0).
\]
Set $\rho=\sqrt{2c_+E}$ and $w_m=e^{-\rho s_m}$. At a site $m\ne0$,
the conductances toward and away from zero are
\[
\begin{array}{c|cc}
 &m>0&m<0\\ \hline
 c_{\rm in}(m)&c_{m-1}&c_m\\
 c_{\rm out}(m)&c_m&c_{m-1}.
\end{array}
\]
Indeed $c_m$ is the coefficient of the edge $(m,m+1)$. Then
\begin{align*}
 \frac{(Qw)_m}{w_m}
 &=E+c_{\rm out}(1-e^{-\rho/c_{\rm out}})
       +c_{\rm in}(1-e^{\rho/c_{\rm in}})\\
 &\leq E-\frac{\rho^2}{2c_{\rm in}}
 \leq E-\frac{\rho^2}{2c_+}=0.
\end{align*}
Here $1-e^{-x}\leq x$ and $1-e^x\leq-x-x^2/2$ for $x\geq0$.
At the source, $(Qw)_0\leq E+2\rho$. Since $w\in\ell^2(\Z)$,
positivity of $Q^{-1}$ applied to the coordinatewise inequality
$Q(w/(E+2\rho))\leq\delta_0$ gives
\[
 (Q^{-1})_{m0}\geq\frac{e^{-\rho s_m}}{E+2\rho}
 \geq cE^{-1/2}e^{-C\sqrt E|m|}.
\]
Here $s_m\le |m|/c_-$ and
$E+2\rho\le(1+2\sqrt{2c_+})\sqrt E$ for $E\le1$.
Translation proves the general case.
\end{proof}

For $E<a/2$, the coefficients in \eqref{tr:eq:schur} obey
$1/b\leq c_m(E)\leq2/a$. Thus, for any fixed $d>0$,
\begin{equation}\label{tr:eq:free-real}
 \begin{aligned}
 |G(2m,0;E)|&\leq CE^{-1/2} &&(m\in\Z),\\
 |G(2m,0;E)|&\geq c_dE^{-1/2} &&(|m|\leq dE^{-1/2}).
 \end{aligned}
\end{equation}
Indeed $G(2m,0;E)=-(-1)^m(Q_E^{-1})_{m0}$. Taking absolute values
removes the sign, and $e^{-C\sqrt E|m|}\ge e^{-Cd}$ on the indicated interval.

\subsection{The active-site channel}

\begin{lemma}\label{tr:lem:gradient}
There are $E_0,c,C>0$, depending only on $a,b$, such that for every
configuration satisfying \eqref{tr:eq:assumption} and every $0<E<E_0$
there is a set $S_E\subset\Z$ such that
\begin{equation}\label{tr:eq:shell}
 |S_E|\geq cE^{-1/2},\qquad
 cE^{-1/2}\leq|m|\leq CE^{-1/2}\quad(m\in S_E),
\end{equation}
and
\begin{equation}\label{tr:eq:active-real}
 |G(2m,1;E)|\geq c\quad(m\in S_E),\qquad
 |G(2m,1;E)|\leq1\quad(m\in\Z).
\end{equation}
The set $S_E$ consists of cell indices on one side of cell zero.
Cell zero contains the input site $1$ in $G(2m,1;E)$; the side may
depend on $E$ and the configuration.
\end{lemma}

\begin{proof}
Let $h_m=(A_E^{-1})_{m0}$ and $\mu_m=E(v_m-E)$. Positivity follows as in
Lemma~\ref{tr:lem:conductance}. To see strict positivity, note that
\[
 ((I-A_E/\Lambda)u)_m
 =\left(1-\frac{2+\mu_m}{\Lambda}\right)u_m
   +\frac{u_{m-1}+u_{m+1}}{\Lambda},\qquad \Lambda>\|A_E\|.
\]
The coefficients are nonnegative. For $d=|i-j|$, the straight path in
the $d$th power gives
$[(I-A_E/\Lambda)^d]_{ij}\ge\Lambda^{-d}$, so
$(A_E^{-1})_{ij}\ge\Lambda^{-(d+1)}>0$ by the Neumann series.
For $E<a/2$,
\[
 \tfrac a2 E\leq\mu_m\leq bE,\qquad
 h_0\leq((L_0+\tfrac a2 E)^{-1})_{00}\leq CE^{-1/2}.
\]
Since $h\in\ell^2$, $h_m\to0$ at both ends. On either half-line away
from zero, its equation is
$h_{m+1}-2h_m+h_{m-1}=\mu_mh_m>0$.
For $m\ge1$, the differences $d_m=h_m-h_{m-1}$ therefore satisfy
$d_{m+1}-d_m=\mu_mh_m>0$. If any $d_m\ge0$, all subsequent
differences after the next step would have a strictly positive lower
bound, contradicting $h_m\to0$. Hence $h_0>h_1>h_2>\cdots>0$;
the same argument applies to $h_{-m}$. In particular $h_m\le h_0$.
Define outgoing edge gradients
\[
 j_k^+=h_k-h_{k+1},\qquad j_k^-=h_{-k}-h_{-k-1}\qquad(k\geq0).
\]
They are positive, decrease with $k$, and tend to zero. The equation at
zero gives
\begin{equation}\label{tr:eq:flux}
 j_0^++j_0^-+\mu_0h_0=1.
\end{equation}
In particular every $j_k^\pm\leq1$. Since $\mu_0h_0\leq C\sqrt E$,
at least one of $j_0^+,j_0^-$ is at least $1/4$ for small $E$.
On that half-line,
\[
 0\leq j_0^\pm-j_k^\pm
    =\sum_{\ell=1}^k\mu_{\pm\ell}h_{\pm\ell}
    \leq Ck\sqrt E.
\]
Choose $d>0$ small enough that $Cd\leq1/8$.
Let $\mathcal I_E$ be the integers $k$ with
$d/(4\sqrt E)\leq k\leq d/(2\sqrt E)$. The chosen gradient is at least
$1/8$ there. Set
\[
 S_E=\{k+1:k\in\mathcal I_E\}\quad\hbox{on the right},\qquad
 S_E=\{-k:k\in\mathcal I_E\}\quad\hbox{on the left}.
\]
For sufficiently small $E$,
$|\mathcal I_E|\ge d/(4\sqrt E)-1\ge d/(8\sqrt E)$.
The same interval gives the distance bounds in \eqref{tr:eq:shell}.
Finally \eqref{tr:eq:blocks} gives
$|G(2m,1;E)|=|h_m-h_{m-1}|$, proving both claims.
\end{proof}

\subsection{Column norms and complex-energy stability}

\begin{lemma}\label{tr:lem:columns}
There are $C,E_0>0$, depending only on $a,b$, such that for every
configuration satisfying \eqref{tr:eq:assumption}, every $j\in\Z$
and every $0<E<E_0$,
\begin{equation}\label{tr:eq:columns}
 \norm{R(E)\delta_{2j}}^2\leq CE^{-3/2},\qquad
 \norm{R(E)\delta_{2j+1}}^2\leq CE^{-1/2}.
\end{equation}
\end{lemma}

\begin{proof}
On the real gap, $R'(E)=R(E)^2$; its diagonal at a source equals the
squared norm of that resolvent column. In the zero-site block,
\[
 \frac d{dE}(-Q_E^{-1})=Q_E^{-1}Q_E'Q_E^{-1},\qquad
 Q_E'=I+D^*c(E)^2D.
\]
Since $c_m(E)^2\leq(2/a)c_m(E)$ for $E<a/2$, quadratic forms satisfy
$Q_E'\leq CE^{-1}Q_E$ when $E\leq1$.
Lemma~\ref{tr:lem:conductance} therefore bounds the relevant diagonal by
\[
 CE^{-1}(Q_E^{-1})_{jj}\leq CE^{-3/2}.
\]

In the active block, $A_E'=V-2E$. For $E<a/4$ this is positive, and
\[
 \frac d{dE}(EA_E^{-1})
 =A_E^{-1}-EA_E^{-1}(V-2E)A_E^{-1}\leq A_E^{-1}.
\]
Its diagonal is nonnegative because it is a diagonal entry of $R(E)^2$.
Comparison with $L_0+(a/2)E$ bounds it above by $CE^{-1/2}$.
All arguments apply at every source index.
\end{proof}

The column bounds control the error when a gap energy acquires a small
imaginary part. Their different powers are needed to preserve the
active-site channel, whose real amplitude is only of order one.
\begin{proposition}\label{tr:prop:complex}
There are $K\geq1$, $E_0>0$, and $c,C>0$, depending only on
$a,b$, with the following property. If
\begin{equation}\label{tr:eq:cone}
 0<E<E_0,\qquad0<\varepsilon\leq E/K,\qquad z=E+i\varepsilon,
\end{equation}
then, for all $m\in\Z$,
\begin{equation}\label{tr:eq:complex-upper}
 |G(2m,0;z)|\leq CE^{-1/2},\qquad |G(2m,1;z)|\leq C.
\end{equation}
There are sets $S_E^{(0)},S_E^{(1)}$, each satisfying the shell bounds
\eqref{tr:eq:shell}, on which respectively
\begin{equation}\label{tr:eq:complex-lower}
 |G(2m,0;z)|\geq cE^{-1/2}\quad(m\in S_E^{(0)}),\qquad
 |G(2m,1;z)|\geq c\quad(m\in S_E^{(1)}).
\end{equation}
The sets can be chosen independently of $\varepsilon$ in \eqref{tr:eq:cone}.
\end{proposition}

\begin{proof}
The spectral theorem gives $\norm{R(E+i\varepsilon)\delta_j}
\leq\norm{R(E)\delta_j}$. The resolvent identity and
Cauchy--Schwarz therefore imply
\begin{equation}\label{tr:eq:resolvent-stability}
 |G(n,j;E+i\varepsilon)-G(n,j;E)|
 \leq\varepsilon\norm{R(E)\delta_n}\norm{R(E)\delta_j}.
\end{equation}
For $n=2m$, Lemma~\ref{tr:lem:columns} gives explicitly
\[
\begin{aligned}
 |G(2m,0;E+i\varepsilon)-G(2m,0;E)|&\le C\varepsilon E^{-3/2},\\
 |G(2m,1;E+i\varepsilon)-G(2m,1;E)|&\le C\varepsilon E^{-1}.
\end{aligned}
\]
For $\varepsilon\le E/K$, these are at most $(C/K)E^{-1/2}$ and
$C/K$, respectively.

Choose for $S_E^{(0)}$, for example, the integers
$E^{-1/2}\leq m\leq2E^{-1/2}$, and use
\eqref{tr:eq:free-real} with $d=2$. Choose $S_E^{(1)}$ as in
Lemma~\ref{tr:lem:gradient}. Taking $K$ sufficiently large preserves half
of each real lower bound. The real global upper bounds and the same error
estimates give \eqref{tr:eq:complex-upper}.
\end{proof}

\begin{proof}[Proof of Theorem~\ref{tr:thm:main}]
Write $r=r_f$, $s=s_f$. On the cone \eqref{tr:eq:cone}, a nonzero polynomial
of vanishing order $k$ at zero has absolute value comparable to $E^k$
for all sufficiently small $E$. The constants may depend on the
polynomial. For small $E$ the shells in Proposition~\ref{tr:prop:complex}
satisfy $\{2m:m\in S_E^{(i)}\}\cap F_f=\varnothing$ for $i=0,1$,
because $|2m|\ge cE^{-1/2}$ whereas $F_f$ is fixed. Thus
\eqref{tr:eq:tail-module} applies at these physical sites.

If $r\leq s$, then $r<\infty$ and $\nu_H(f)=4r+1$. On $S_E^{(0)}$,
\begin{align*}
 |(R(z)f)(2m)|
 &\geq |A_f(z)|\,|G(2m,0;z)|
            -|B_f(z)|\,|G(2m,1;z)|\\
 &\geq cE^{r-1/2}-CE^s\geq c' E^{r-1/2}
\end{align*}
for small $E$, with the second term absent if $B_f=0$.
If $s<r$, then $s<\infty$, $r\geq s+1$ unless $A_f=0$, and
$\nu_H(f)=4s+3$. On $S_E^{(1)}$ the reverse triangle inequality gives
\[
 |(R(z)f)(2m)|\geq cE^s-CE^{r-1/2}\geq c'E^s,
\]
with the second term absent if $A_f=0$.
Thus the two channel amplitudes cannot cancel at leading order: their
possible powers differ by a half-integer, while $r,s$ are integers.

Each shell has at least $cE^{-1/2}$ sites, all with $|2m|\geq cE^{-1/2}$.
In both cases, it follows that
\begin{equation}\label{tr:eq:weighted-resolvent}
 \sum_{n\in\Z}|n|^q|(R(E+i\varepsilon)f)(n)|^2
 \geq c E^{(\nu_H(f)-q)/2-2}.
\end{equation}
Take $\varepsilon=T^{-1}$ and integrate over $E\in[K/T,2K/T]$.
Here $\varepsilon\le E/K$, and $T>2K/E_0$ ensures \eqref{tr:eq:cone}.
Increase $T$ further, depending on $f$, for the polynomial bounds and
the disjointness of the shells from $F_f$.
By \eqref{tr:eq:parseval},
\[
 \M_q(T;f)\geq\frac cT\int_{K/T}^{2K/T}
          E^{(\nu_H(f)-q)/2-2}\,dE
       \geq c' T^{(q-\nu_H(f))/2}.
\]
The last assertion of the theorem follows by logarithms and a lower limit.
\end{proof}

\begin{corollary}\label{tr:cor:sites}
Under \eqref{tr:eq:assumption}, for each fixed $j\in\Z$ and $q>0$,
\begin{equation}\label{tr:eq:point-sites}
 \M_q(T;\delta_{2j})\geq cT^{(q-1)/2},\qquad
 \M_q(T;\delta_{2j+1})\geq cT^{(q-3)/2}
\end{equation}
for all sufficiently large $T$. For fixed $a,b,q,j$, constants can be chosen
uniformly over configurations satisfying \eqref{tr:eq:assumption}.
In particular,
\[
 \M_2(T;\delta_{2j})\geq cT^{1/2},\qquad
 \M_4(T;\delta_{2j+1})\geq cT^{1/2}.
\]
No growing second-moment lower bound for $\delta_{2j+1}$ follows from this
corollary.
\end{corollary}
\begin{proof}[Proof of Corollary~\ref{tr:cor:sites}]
For $\delta_0$ and $\delta_1$ the channel pairs are $(1,0)$ and $(0,1)$,
so their loss indices are $1$ and $3$. Translation by an even number of
sites preserves \eqref{eq:mosaic-operator}--\eqref{tr:eq:assumption}; the Green estimates
above are uniform over all such translated configurations. On a shell
escaping a fixed source $2j$, the weight $|n|^q$ is comparable to
$|n-2j|^q$. This proves \eqref{tr:eq:point-sites}, including uniformity.

Equivalently, write $\delta_n=A_n(H)\delta_0+B_n(H)\delta_1$.
Evaluation of the recurrence at zero gives
\[
 A_{2j}(0)=(-1)^j,\qquad A_{2j+1}(0)=0,\qquad
 B_{2j+1}(0)=(-1)^j.
\]
Hence \eqref{tr:eq:index} gives the same loss indices at every site.
\end{proof}

\subsection{Examples and the two channel scales}
The channel polynomials make higher-order threshold cancellations explicit.
For example, for integers $k\geq0$,
\begin{equation}\label{tr:eq:examples}
 \begin{aligned}
 \nu_H(H^k\delta_0)&=4k+1,&
 \nu_H(H^k\delta_1)&=4k+3,\\
 \nu_H(\delta_{-1}+\delta_1)&=5,&
 \nu_H(\delta_{-1}-\delta_1)&=3.
 \end{aligned}
\end{equation}
The last two identities follow from $H\delta_0=\delta_{-1}+\delta_1$.
These identities hold for every bounded positive configuration. Under
nondegenerate iid randomness, Theorem~\ref{up:thm:front} converts them
into the exact transport exponents; without randomness they give only
the lower bounds of Theorem~\ref{tr:thm:main}.

The elementary scale bookkeeping is useful. At $E\asymp T^{-1}$ in the
gap, the contributing distance is $E^{-1/2}\asymp\sqrt T$.
The zero-site Green amplitude is $E^{-1/2}$, whereas the active-site
amplitude on the even output sublattice is of order one. The squared
amplitude, the number of sites, the energy-window width, and the prefactor
in Parseval account for the two losses $1$ and $3$. A zero of order $k$
in a dominant polynomial channel contributes $E^{2k}$ and raises the
loss index by $4k$.

\section{Random cancellation and physical resolvent mass}
\label{sec:random-transport}
We prove Theorem~\ref{rt:thm:main} under \eqref{rt:eq:law}.
Write $v_m=v_\omega(m)$ and $R(z)=(H_\omega-z)^{-1}$.
Unlike the gap-side argument, we work at negative real energies. The
main estimate is Lemma~\ref{rt:lem:complex}: on a good transfer event,
it bounds the physical mass on two distant shells while retaining both
source amplitudes. Lemma~\ref{rt:lem:energy} then controls the measure
of good energies. The polynomial tail identity and Parseval formula
in Lemmas~\ref{tr:lem:module} and~\ref{tr:lem:parseval} convert these
estimates into the moment lower bound.

\subsection{The scaled source jump}
In the staggered coordinates of Section~\ref{sec:gap-schur},
for real $\kappa>0$ and $z=-\kappa^2+i\varepsilon$ define
\[
 w_m=\begin{pmatrix}\psi_m\\ \sqrt{\bar v}\,\chi_{m-1}/\kappa\end{pmatrix}.
\]
Away from a source, $(H-z)u=0$ gives
\begin{equation}\label{rt:eq:cell}
 w_{m+1}=C_{v_m}(z;\kappa)w_m,\qquad
 C_v(z;\kappa)=
 \begin{pmatrix}
 1+z(v-z)&\kappa(v-z)/\sqrt{\bar v}\\
 \sqrt{\bar v}\,z/\kappa&1
 \end{pmatrix}.
\end{equation}
Its determinant is exactly one. We hold the real scaling $\kappa$ fixed
when changing the imaginary part of $z$.

Let $u=R(z)(\alpha\delta_0+\beta\delta_1)$, with arbitrary
$\alpha,\beta\in\C$. The inhomogeneous equations at cell zero are
\[
 \chi_0-\chi_{-1}-z\psi_0=\alpha,\qquad
 \psi_0-\psi_1+(v_0-z)\chi_0=\beta.
\]
Thus
\begin{equation}\label{rt:eq:jump}
 w_1=C_{v_0}(z;\kappa)w_0+b,\qquad
 b=\begin{pmatrix}(v_0-z)\alpha-\beta\\
                  \sqrt{\bar v}\,\alpha/\kappa\end{pmatrix}.
\end{equation}
For sufficiently small $\kappa$ and $0\le\varepsilon\le\kappa^2$, uniformly in
$v_0\in[a,b]$,
\begin{equation}\label{rt:eq:source-lower}
 \norm{w_0}^2+\norm{w_1}^2
 \ge c\norm b^2
 \ge c\bigl(\abs{\alpha}^2/\kappa^2+\abs{\beta}^2\bigr).
\end{equation}
Indeed $\norm{C_{v_0}(z;\kappa)}\le C$,
$\alpha=\kappa b_2/\sqrt{\bar v}$, and
$\beta=(v_0-z)\kappa b_2/\sqrt{\bar v}-b_1$.
This handles both half-lines simultaneously, without choosing a
favourable Weyl function or assuming independence of a boundary value.

\subsection{Maximal products}
At real energy, set $C_v(\kappa)=C_v(-\kappa^2;\kappa)$ and
$C_\kappa=\E C_{v_0}(\kappa)$.
They have the exact affine decomposition
\begin{equation}\label{rt:eq:affine}
 C_v(\kappa)=C_\kappa+(v-\bar v)D_\kappa,\qquad
 D_\kappa=\begin{pmatrix}-\kappa^2&\kappa/\sqrt{\bar v}\\0&0\end{pmatrix},
\end{equation}
where
\begin{equation}\label{rt:eq:mean}
 C_\kappa=
 \begin{pmatrix}
 1-\bar v\kappa^2-\kappa^4&
       \kappa\sqrt{\bar v}+\kappa^3/\sqrt{\bar v}\\
 -\kappa\sqrt{\bar v}&1
 \end{pmatrix}.
\end{equation}
In particular $\det C_\kappa=1$, even though averaging generally does not
preserve the determinant.

Put $J=\left(\begin{smallmatrix}0&-1\\1&0\end{smallmatrix}\right)$ and
$R_t=e^{tJ}$. For small $\kappa$ define $\theta_\kappa\in(0,\pi)$ by
\[
 \cos\theta_\kappa=1-\tfrac12(\bar v\kappa^2+\kappa^4).
\]
Then $\theta_\kappa=\sqrt{\bar v}\kappa+O(\kappa^3)$. Cayley--Hamilton gives
$K_\kappa^2=-I$ for
$K_\kappa=(C_\kappa-\cos\theta_\kappa I)/\sin\theta_\kappa$.
Thus $C_\kappa=\cos\theta_\kappa I+\sin\theta_\kappa K_\kappa$,
and multiplication, also for negative powers, gives
\begin{equation}\label{rt:eq:mean-powers}
 C_\kappa^n=\cos(n\theta_\kappa)I+
   \sin(n\theta_\kappa)
       \frac{C_\kappa-\cos\theta_\kappa I}{\sin\theta_\kappa},
 \qquad n\in\Z.
\end{equation}
The quotient is $-J+O(\kappa)$. Since $|\sin(n\theta_\kappa)|\le1$,
the error does not accumulate with $n$. Consequently, uniformly for every
integer $n$,
\begin{equation}\label{rt:eq:rotation-close}
 \norm{C_\kappa^n-R_{-n\theta_\kappa}}\le C\kappa,\qquad
 \norm{C_\kappa^n}+\norm{C_\kappa^{-n}}\le C.
\end{equation}

For products starting immediately to the right and left of the source,
write
\[
 \begin{aligned}
 F_n^+(\kappa)&=C_{v_n}(\kappa)\cdots C_{v_1}(\kappa),\\
 F_n^-(\kappa)&=C_{v_{-n}}(\kappa)^{-1}\cdots C_{v_{-1}}(\kappa)^{-1},
 \qquad F_0^\pm=I.
 \end{aligned}
\]
\begin{lemma}\label{rt:lem:maximal}
For every fixed $\delta>0$, all sufficiently small $\kappa$, and all
$N\ge1$,
\begin{equation}\label{rt:eq:bad-prob}
 \Prob\left(
 \max_{\substack{0\le n\le N\\\sigma\in\{+,-\}}}
 \norm{F_n^\sigma(\kappa)-C_\kappa^{\sigma n}}>\delta
 \right)
 \le C_\delta\exp\left(-\frac{c_\delta}{N\kappa^2}\right).
\end{equation}
Here $C_\kappa^{+n}=C_\kappa^n$ and
$C_\kappa^{-n}$ has its usual meaning.
\end{lemma}
\begin{proof}
Consider the plus products and put $Y_n=C_\kappa^{-n}F_n^+$.
With $\xi_{n+1}=v_{n+1}-\bar v$,
\[
 Y_{n+1}=(I+\xi_{n+1}K_n)Y_n,\qquad
 K_n=C_\kappa^{-n-1}D_\kappa C_\kappa^n,\qquad
 \norm{K_n}\le C\kappa.
\]
Thus $Y_n$ is a real matrix-valued martingale with $Y_0=I$.
For a fixed $r>0$, stop it at
$\tau=\inf\{n:\norm{Y_n-I}_F>r\}$, where $\norm\cdot_F$ denotes
the Frobenius norm. Before exit, $\norm{Y_n}_F\le\sqrt2+r$, so each
entry of $Y_{n\wedge\tau}-I$ has centered increments bounded by
$C_r\kappa$. If $\tau\le N$, one of its four entries has an absolute
value greater than $r/2$ at some time at most $N$.
The scalar maximal exponential estimate \cite[Lemma~2]{Azuma}, followed
by a union bound over the four entries, gives
\[
 \Prob\left(\max_{n\le N}\norm{Y_n-I}_F>r\right)
 \le 32\exp\left(-\frac{c_r}{N\kappa^2}\right).
\]
Only the stopped increments have been bounded; no bound on an
unrestricted product is assumed. Since
$F_n^+-C_\kappa^n=C_\kappa^n(Y_n-I)$ and the mean powers are uniformly
bounded, choose $r$ proportional to $\delta$ to obtain the plus estimate.

Because the adjugate is linear for $2$ by $2$ matrices and each
$C_v(\kappa)$ has determinant one,
\[
 \E C_{v_0}(\kappa)^{-1}=C_\kappa^{-1},\qquad
 C_v(\kappa)^{-1}
 =C_\kappa^{-1}+(v-\bar v)
       \begin{pmatrix}0&-\kappa/\sqrt{\bar v}\\0&-\kappa^2\end{pmatrix}.
\]
The variables $v_{-1},v_{-2},\ldots$ are iid. Applying the same stopping
argument gives the minus estimate. A union bound proves
\eqref{rt:eq:bad-prob}; independence between the two good events is unnecessary.
\end{proof}

\subsection{Complex stability and observation of the physical component}
Fix a small numerical $\delta_0>0$, to be chosen below.
Call a real energy $E=-\kappa^2$ \emph{good at length $N$} if the maximum
inside \eqref{rt:eq:bad-prob} is at most $\delta_0$.
The good event is measurable in $(\omega,E)$, since it is defined by
finitely many continuous matrix products.

\begin{lemma}\label{rt:lem:complex}
There are constants $\rho_0>0$, $K_0<\infty$ and $\kappa_0>0$,
depending only on $a,b,\bar v$, such that the following holds.
Suppose $0<\kappa<\kappa_0$, $E=-\kappa^2$ is good at length $N$,
$0<\varepsilon\le\kappa^2$, and
\begin{equation}\label{rt:eq:scales}
 N\varepsilon/\kappa\le\rho_0,\qquad N\kappa\ge K_0.
\end{equation}
For $z=E+i\varepsilon$ and arbitrary $\alpha,\beta\in\C$,
\begin{equation}\label{rt:eq:shell-bridge}
 \sum_{N/2\le |m|\le N}
 \abs{\bigl(R(z)(\alpha\delta_0+\beta\delta_1)\bigr)(2m)}^2
 \ge cN\bigl(\abs{\alpha}^2/\kappa^2+\abs{\beta}^2\bigr).
\end{equation}
\end{lemma}
\begin{proof}
First keep the real normalization $\kappa$ fixed. Directly from
\eqref{rt:eq:cell}, uniformly in $v\in[a,b]$ and $\varepsilon\le\kappa^2$,
\begin{equation}\label{rt:eq:single-complex}
 \norm{C_v(E+i\varepsilon;\kappa)-C_v(E;\kappa)}\le C\varepsilon/\kappa.
\end{equation}
The inverse matrices satisfy the same estimate by their explicit adjugates.
On the good event all real partial products have uniformly bounded norms
and inverses, by \eqref{rt:eq:rotation-close} and determinant one.
Take either $A_j=C_{v_j}(E;\kappa)$ for every $j\ge1$, or
$A_j=C_{v_{-j}}(E;\kappa)^{-1}$ for every $j\ge1$.
Let $\widetilde A_j$ be their complex counterparts and put
$P_n=A_n\cdots A_1$, $\widetilde P_n=\widetilde A_n\cdots\widetilde A_1$.
Thus $P_n$ is respectively $F_n^+(\kappa)$ or $F_n^-(\kappa)$.
Variation of constants is the exact identity
\[
 \widetilde P_n-P_n
 =\sum_{j=1}^n P_nP_j^{-1}
              (\widetilde A_j-A_j)\widetilde P_{j-1}.
\]
For completeness, put $d=\varepsilon/\kappa$. Choose $B\ge1$ so that
$\|P_n\|,\|P_n^{-1}\|\le B$ for $0\le n\le N$, and $L$ so that
$\|\widetilde A_j-A_j\|\le Ld$. With $c_*=B^2L$ and
$a_n=\|\widetilde P_n\|$, the identity implies
$a_n\le B+c_*d\sum_{j=0}^{n-1}a_j$. Induction yields
$a_n\le B(1+c_*d)^n$, and substitution into the same identity gives
\[
 \|\widetilde P_n-P_n\|
 \le B\bigl((1+c_*d)^n-1\bigr)
 \le B(e^{c_*nd}-1)\le Bc_*nd\,e^{c_*nd}.
\]
In particular, the difference tends to zero with $Nd$, and
\begin{equation}\label{rt:eq:complex-products}
 \max_{n\le N}\norm{\widetilde P_n-P_n}
 \le C(N\varepsilon/\kappa)\exp(CN\varepsilon/\kappa).
\end{equation}
This applies on both half-lines. Choose $\delta_0$, then $\rho_0$, then
$\kappa_0$ so small that \eqref{rt:eq:rotation-close} and
\eqref{rt:eq:complex-products} ensure that the complex plus and minus products
are within $d_0=1/16$ of $R_{-n\theta_\kappa}$ and
$R_{n\theta_\kappa}$, respectively, uniformly for $n\le N$.

Write $e_1=(1,0)^t$. For an interval $I$ of $M$ consecutive integers,
put $S=\sum_{n\in I}e^{2in\theta}$. Direct multiplication gives
\[
 \sum_{n\in I}R_{n\theta}^t e_1e_1^t R_{n\theta}
 =\frac M2I+\frac12
 \begin{pmatrix}\Re S&-\Im S\\-\Im S&-\Re S\end{pmatrix}.
\]
Its eigenvalues are $(M\pm|S|)/2$, and the geometric sum satisfies
$|S|\le1/|\sin\theta|$. Thus for any $x\in\C^2$ and
either sign of $\theta=\theta_\kappa$,
\begin{equation}\label{rt:eq:gramian}
 \sum_{n\in I}\abs{e_1^tR_{n\theta}x}^2
 \ge \left(\frac M2-\frac1{2|\sin\theta|}\right)\norm x^2.
\end{equation}
This is the least-eigenvalue bound for
$\sum R_{n\theta}^t e_1e_1^t R_{n\theta}$ and remains valid for complex
vectors. If matrices $P_n$ are within $d_0$ of these rotations, then
$|a+b|^2\ge |a|^2/2-|b|^2$ implies
\[
 \sum_{n\in I}\abs{e_1^tP_nx}^2
 \ge\left(\frac M4-\frac1{4|\sin\theta|}-Md_0^2\right)\norm x^2.
\]
For intervals of length comparable to $N$, the right side is at least
$cN\norm x^2$ when $K_0$ in \eqref{rt:eq:scales} is large enough.

For the actual resolvent solution, on the right
$w_m=F_{m-1}^+(z;\kappa)w_1$ for $m\ge1$; on the left
$w_{-k}=F_k^-(z;\kappa)w_0$ for $k\ge1$.
Apply the preceding observation bound to each shell
$\lceil N/2\rceil\le m,k\le N$.
Since $u(2m)=(-1)^m e_1^tw_m$, their sum is bounded below by
$cN(\norm{w_0}^2+\norm{w_1}^2)$.
Finally \eqref{rt:eq:source-lower} proves \eqref{rt:eq:shell-bridge}.
\end{proof}

\begin{remark}
A pointwise comparison of the physical vector
$(\psi,\chi)$ with $(\psi,\sqrt{\bar v}\chi/\kappa)$ can lose a factor
$\kappa^2$ in squared norms. Equation \eqref{rt:eq:gramian} avoids that
loss: over many rotations the unscaled physical component $\psi$ sees a
fixed fraction of the entire scaled norm. The source jump then supplies
$\kappa^{-2}$ for the zero-site channel and $1$ for the active-site
channel. The estimate is deterministic on the simultaneous good event.
For deterministic $\alpha,\beta$, taking expectation gives the lower bound
$cN(|\alpha|^2/\kappa^2+|\beta|^2)$ times the probability of the good
event. This operation does not extract sample-dependent channel
polynomials from an expectation; such polynomials are used pathwise
in part~\textup{(i)} below.
\end{remark}

\subsection{Energy integration}
Choose a sufficiently small fixed $c_0>0$ with $c_0<\rho_0$.
For a scale $h=h(T)$ tending to zero set
\begin{equation}\label{rt:eq:window}
 I_T=[-2h^2,-h^2],\qquad
 N_T=\lfloor c_0hT\rfloor,\qquad \varepsilon=T^{-1}.
\end{equation}
For $E\in I_T$ put $\kappa(E)=\sqrt{-E}\in[h,\sqrt2h]$.
As soon as $Th^2\to\infty$, we have $\varepsilon\le\kappa(E)^2$,
$N_T\kappa(E)\to\infty$, and
$N_T\varepsilon/\kappa(E)\le c_0$, uniformly on the window.
The deterministic bridge therefore applies to all its good energies.

\begin{lemma}\label{rt:lem:energy}
Let $B_T(\omega)$ be the Lebesgue measure of the energies in $I_T$
which are not good at length $N_T$. If $N_Th^2\le1$, then
\begin{equation}\label{rt:eq:bad-measure}
 \E B_T\le C h^2\exp\left(-\frac{c}{N_Th^2}\right),\qquad
 \Prob(B_T>h^2/2)\le C\exp\left(-\frac{c}{N_Th^2}\right).
\end{equation}
\end{lemma}
\begin{proof}
For each fixed $E\in I_T$, Lemma~\ref{rt:lem:maximal} bounds its failure
probability by $C\exp(-c/(N_Th^2))$, uniformly on the window.
Tonelli's theorem and $|I_T|=h^2$ give the first bound.
Markov's inequality gives the second. This does not require independence
between different energies, or a uniform good event at every energy.
\end{proof}

\begin{proof}[Proof of Theorem~\ref{rt:thm:main}\textup{(i)}]
Choose a sufficiently small deterministic $d>0$ and take
\begin{equation}\label{rt:eq:h}
 h(T)=dT^{-1/3}(\log\log T)^{-1/3},\qquad T>e^e.
\end{equation}
Then $Th^2\to\infty$ and $N_Th^2\le c_0d^3/(\log\log T)$.
Choose $d$ so small that \eqref{rt:eq:bad-measure} is bounded by
$C(\log T)^{-3}$. At dyadic times $T_j=2^j$ these probabilities are
summable. By Borel--Cantelli, on an event of
probability one, for every sufficiently large $j$ the good subset
$G_{T_j}\subset I_{T_j}$ has measure at least $h(T_j)^2/2$.
This event does not depend on $q$ or $f$.

Fix such a sample and a nonzero finitely supported $f$. Its channel
polynomials are now fixed. If a nonzero polynomial has order $r$ at zero,
its magnitude is bounded below by a positive constant times $h^{2r}$
at every $z=E+i/T$ with $E\in I_T$, for all sufficiently large $T$.
This follows by factoring off $z^r$, since $|z|\asymp h^2$.
Thus, writing $\nu=\nu_H(f)$,
\begin{equation}\label{rt:eq:jet-weight}
 \frac{|A_f(z)|^2}{\kappa(E)^2}+|B_f(z)|^2
 \ge c_f h^{\nu-3}.
\end{equation}
Zero polynomials simply contribute zero; at least one is nonzero.
The exponent is
$\min(4r_f-2,4s_f)=\nu-3$.
For large $T$, the shells in Lemma~\ref{rt:lem:complex} are outside the
finite set $F_f$ in Lemma~\ref{tr:lem:module}. Since $|2m|\ge N_T$ on these
shells, \eqref{tr:eq:tail-module}, \eqref{rt:eq:shell-bridge} and
\eqref{rt:eq:jet-weight} give, on every good energy,
\[
 \sum_n |n|^q |(R(E+i/T)f)(n)|^2
 \ge c_f N_T^{q+1}h^{\nu-3}.
\]
At the large dyadic times integrate over $G_T$ and use Parseval:
\begin{align}\label{rt:eq:final-scaling}
 \M_q(T;f)
 &\ge \frac{c_f}{T}N_T^{q+1}h^{\nu-3}h^2
 \ge c_f T^q h^{q+\nu}\notag\\
 &=c_f T^{(2q-\nu)/3}
                (\log\log T)^{-(q+\nu)/3}.
\end{align}

To pass to every large time, for $T_j\le T\le2T_j$ positivity of the
integrand in \eqref{tr:eq:moment} gives
\begin{equation}\label{rt:eq:time-comparison}
 \M_q(T;f)\ge(T_j/T)\M_q(T_j;f)\ge\tfrac12\M_q(T_j;f).
\end{equation}
Powers and double logarithms of $T_j$ and $T$ are comparable for every
fixed $q,\nu$, including when the power is negative.
This proves \eqref{rt:eq:quenched}. All estimates on this event apply to
every finite state and every $q>0$, with state-dependent constants; no
uncountable intersection is needed. Taking logarithms gives
$\beta_f^-(q)\ge 2/3-\nu_H(f)/(3q)$.
\end{proof}

\begin{proof}[Proof of Theorem~\ref{rt:thm:main}\textup{(ii)}]
Use \eqref{rt:eq:window} with $h=dT^{-1/3}$, where the fixed $d>0$ is
sufficiently small. Then $N_Th^2\le c_0d^3$, so
Lemma~\ref{rt:lem:maximal} makes the probability of a good energy at least
$1/2$, uniformly for every $E\in I_T$ and every large $T$.
For $\delta_0$, \eqref{rt:eq:shell-bridge} gives the weighted lower bound
$cN_T^{q+1}h^{-2}$ on the good event; for $\delta_1$ it gives
$cN_T^{q+1}$. These constants are deterministic, regardless of the
random boundary values of the actual resolvent. Taking expectation of
these pointwise lower bounds and using Tonelli and Parseval yields
\[
 \E\M_q(T;\delta_0)\ge cT^{-1}N_T^{q+1},\qquad
 \E\M_q(T;\delta_1)\ge cT^{-1}N_T^{q+1}h^2.
\]
As $N_T\asymp hT$, these are exactly the two powers in
\eqref{rt:eq:annealed}. Translation by $2j$ preserves the law and the two
sublattices. On the translated far shell, $|n|$ and $|n-2j|$ are
comparable for each fixed $j$ and all sufficiently large $T$.
This proves the statement at every indicated site.
\end{proof}

The scale balance is distinct from that of the gap-side proof.
There the distance is of order $T^{1/2}$ and no averaging is available.
Here the complex-energy error is small on distances $N\asymp hT$,
while the centered random-product error has size controlled by $Nh^2$.
For disorder averages a small fixed value of $Th^3$ suffices. Almost
surely, the double-logarithmic reduction of $h$ makes the exponential
bad-energy bounds summable on dyadic times. This loss is not asserted
to be necessary.

\section{Growth of random transfer products}
\label{sec:transfer-growth}
We now prove the transfer-growth estimate used in the upper bound.
Assume \eqref{rt:eq:law}. The normalization and local expansions below
also hold for a constant law; nondegeneracy is imposed when deriving
growth. We first work near zero energy and then treat compact energy
sets disjoint from zero.

\subsection{The parabolic normal form}
\label{sec:normal}
For the iid law \eqref{rt:eq:law}, put $\bar v=\E v_0$. For $E=-\kappa^2$ set
\begin{equation}
\label{eq:S-kappa-lyap}
 S_\kappa=\matr{1}{0}{0}{\kappa/\sqrt{\bar v}},\qquad
 B(\kappa,x)=S_\kappa\{-M(-\kappa^2,x)\}S_\kappa^{-1}.
\end{equation}
The exact identity below gives the expansion used in the product estimates.
\begin{lemma}
\label{lem:exact-normal}
For $0<\kappa\le1$ and $x\in[a,b]$,
\begin{equation}
\label{eq:exact-normal}  B(\kappa,x)=
\begin{pmatrix}  1-x\kappa^2-\kappa^4 & -\kappa\sqrt{\bar v}\\[2pt]  \kappa x/\sqrt{\bar v}+\kappa^3/\sqrt{\bar v} & 1
\end{pmatrix}
.
\end{equation}
Consequently, uniformly in $x\in[a,b]$,
\begin{equation}
\label{eq:normal-expansion}  B(\kappa,x)=I+\kappa\sqrt{\bar v}J  +\kappa\frac{x-\bar v}{\sqrt{\bar v}}E_{21}  +\kappa^2R(\kappa,x),
\end{equation}
where $R$ is uniformly bounded for $0\le\kappa\le1$.
More precisely, $R$ is affine in $x$ and satisfies
\begin{equation}
\label{eq:R-Lipschitz}  \sup_{0\le\kappa\le1}\sup_{x\in[a,b]}\norm{R(\kappa,x)}  +\sup_{0\le\kappa\le1}\sup_{x\ne y}\frac{\norm{R(\kappa,x)-R(\kappa,y)}}{|x-y|}<\infty .
\end{equation}
\end{lemma}
\begin{proof}
Direct conjugation gives \eqref{eq:exact-normal}.  Subtracting the two
first-order terms in \eqref{eq:normal-expansion} leaves
\begin{equation*}  R(\kappa,x)=
\begin{pmatrix}  -x-\kappa^2 & 0\\[2pt]  \kappa/\sqrt{\bar v} & 0
\end{pmatrix}
.
\end{equation*}
It is affine in $x$ and satisfies \eqref{eq:R-Lipschitz}.
\end{proof}
\subsection{Phase and radial expansions}
Write $u_\theta=(\cos\theta,\sin\theta)^t$ and
\[
 C_1(x)=\matr{0}{-\sqrt{\bar v}}{x/\sqrt{\bar v}}{0}.
\]
Then $B(\kappa,x)=I+\kappa C_1(x)+\kappa^2R(\kappa,x)$.
The following local calculation is the input to the phase corrector;
it uses only \eqref{rt:eq:law}, not interval support or positive variance.

\begin{lemma}\label{tr:lem:local-expansions}
Define the angle increment by
\[
 B(\kappa,x)u_\theta/\|B(\kappa,x)u_\theta\|=u_{\theta+\delta\theta},
\]
choosing the real angle continuously from
$\delta\theta=0$ at $\kappa=0$. Then
\begin{equation}\label{eq:phase-increment}
 \delta\theta(\theta,x)=
 \kappa\left(\sqrt{\bar v}+
       \frac{x-\bar v}{\sqrt{\bar v}}\cos^2\theta\right)
 +O(\kappa^2).
\end{equation}
Moreover,
\begin{equation}\label{eq:integrand-expansion}
 \log\|B(\kappa,x)u_\theta\|
 =\kappa\langle u_\theta,C_1(x)u_\theta\rangle
   +\kappa^2q(\theta,x)+O(\kappa^3),
\end{equation}
where
\[
 q(\theta,x)=\tfrac12\|C_1(x)u_\theta\|^2+
 \langle u_\theta,R(0,x)u_\theta\rangle
 -\langle u_\theta,C_1(x)u_\theta\rangle^2.
\]
All remainders are uniform in $x\in[a,b]$ and $\theta\in\R$.
The first-order radial term has zero expectation for every $\theta$, and
\begin{equation}\label{eq:Q-fourier}
 Q(\theta):=\E q(\theta,v_0)
 =\frac{\sigma_v^2}{8\bar v}
  +\left(-\frac{\bar v}{2}+\frac{\sigma_v^2}{4\bar v}\right)\cos2\theta
  +\frac{\sigma_v^2}{8\bar v}\cos4\theta.
\end{equation}
\end{lemma}
\begin{proof}
For a real unit vector $u$, the exact signed-angle identity is
\[
 \sin\delta\theta=\frac{\det(u,B(\kappa,x)u)}{\|B(\kappa,x)u\|}
 =\kappa\det(u,C_1(x)u)+O(\kappa^2).
\]
Since $\delta\theta=O(\kappa)$, this also holds with $\delta\theta$
in place of its sine, proving \eqref{eq:phase-increment}.
Expanding the logarithm of the squared norm
and using $R(\kappa,x)=R(0,x)+O(\kappa)$ gives
\eqref{eq:integrand-expansion}.
With $c=\cos\theta$, $s=\sin\theta$,
\[
 \begin{aligned}
 \langle u_\theta,C_1(x)u_\theta\rangle
   &=sc\,\frac{x-\bar v}{\sqrt{\bar v}},\\
 \|C_1(x)u_\theta\|^2&=\bar v s^2+\frac{x^2}{\bar v}c^2,\\
 \langle u_\theta,R(0,x)u_\theta\rangle&=-xc^2.
 \end{aligned}
\]
Taking expectations yields
\[
 Q(\theta)=-\tfrac12\bar v\cos2\theta+
  \frac{\sigma_v^2}{\bar v}c^2\left(\tfrac12-s^2\right),
\]
whose double-angle expansion is \eqref{eq:Q-fourier}.
Boundedness of $x$ and compactness of the angle variable give the
stated uniformity.
\end{proof}
\subsection{Noncompactness away from zero}
We use the Lyapunov exponent per two-site cell: for each fixed $E$,
the following limit exists almost surely and is deterministic,
\begin{equation}\label{eq:block-le}
 \Lambda(E)=\lim_{L\to\infty}\frac1L
 \log\|M(E,v_{L-1})\cdots M(E,v_0)\|.
\end{equation}
Its existence and non-randomness follow from the subadditive ergodic theorem.
For $E\ne0$, the cell matrices are conjugate to
\begin{equation}\label{eq:anderson-matrix}
 \mathcal A(E,x)=\matr{E^2-2-Ex}{-1}{1}{0}.
\end{equation}
Indeed, with $\mathcal C_E=\matr{1}{0}{-1}{E}$,
$\mathcal C_EM(E,x)\mathcal C_E^{-1}=\mathcal A(E,x)$.
The matrices $\mathcal C_E$ and $\mathcal C_E^{-1}$ are uniformly bounded on compact energy
sets disjoint from zero.
\begin{lemma}
\label{lem:furstenberg}
Assume that $\mu$ is not a point mass and let $E\ne0$.
The closed group generated by $\{\mathcal A(E,x):x\in\supp\mu\}$ is noncompact and strongly irreducible on $\R^2$.
Consequently
\begin{equation}
\label{eq:positive-le}  \Lambda(E)>0.
\end{equation}
\end{lemma}
\begin{proof}
Choose $x_1\ne x_2$ in $\supp\mu$.  The generated group $G$ contains
\begin{equation}
\label{eq:commutator-difference}  U:=\mathcal A(E,x_1)\mathcal A(E,x_2)^{-1}  =\matr{1}{t}{0}{1}\in G,  \qquad  t:=E(x_2-x_1)\ne0 .
\end{equation}
The powers of $U$ are unbounded.  If a nonempty finite $G$-invariant set $F$
has cardinality $k$, then $U$ permutes its $k$ lines. Each cycle length
divides $k!$, so $U^{k!}$ fixes every line in $F$. The unique fixed line of
this nontrivial unipotent is $[e_1]$.  Thus $F=\{[e_1]\}$, contradicting
$\mathcal A(E,x)e_1=(E^2-2-Ex,1)^t\notin[e_1]$.  Furstenberg's theorem
\cite{Furstenberg} now gives \eqref{eq:positive-le}.
\end{proof}

\subsection{Finite-length growth}
Here and in Section~\ref{sec:upper-transport}, assume $\sigma_v^2>0$.
Constants may depend on the law and need not be uniform as the variance
tends to zero.
Write
\[
 \mathcal P_N(E)=M(E,v_{N-1})\cdots M(E,v_0).
\]
\begin{lemma}\label{up:lem:products}
There are $\kappa_0,c,C>0$ such that
\begin{equation}\label{up:eq:near-products}
 \E\norm{\mathcal P_N(-\kappa^2)}^{-2}
 \le C\kappa^{-2}e^{-cN\kappa^2},
 \qquad 0<\kappa<\kappa_0,\quad N\ge1.
\end{equation}
For every compact $K_0\subset\R\setminus\{0\}$, there are $c,C>0$ such that
\begin{equation}\label{up:eq:away-products}
 \sup_{E\in K_0}\E\norm{\mathcal P_N(E)}^{-2}\le Ce^{-cN}.
\end{equation}
The same bounds hold for products of inverse cell matrices in the
opposite direction.
\end{lemma}
\begin{proof}
Put $r=\sqrt{\bar v}$, $d=\sigma_v^2/(8\bar v)>0$, and
$A=-\bar v/2+\sigma_v^2/(4\bar v)$. The uniform expansions
\eqref{eq:phase-increment} and \eqref{eq:integrand-expansion} use only
boundedness, the mean, and the variance of the law, so apply under the
present assumptions. With
\[
 Q(\theta)=d+A\cos2\theta+d\cos4\theta,\qquad
 h(\theta)=-\frac{A}{2r}\sin2\theta-\frac{d}{4r}\sin4\theta,
\]
one has $rh'=d-Q$. Let $\theta'$ be the phase of
$B(\kappa,v)u_\theta$ and $\ell=\log\norm{B(\kappa,v)u_\theta}$.
The phase lift obeys $|\theta'-\theta|\le C\kappa$ and
$\E(\theta'-\theta\mid\theta)=\kappa r+O(\kappa^2)$.
Taylor's formula for $h$, whose second derivative is bounded, therefore
gives, uniformly in $\theta$,
\[
 \E\ell=\kappa^2Q(\theta)+O(\kappa^3),\qquad
 \E\{h(\theta')-h(\theta)\}=\kappa r h'(\theta)+O(\kappa^2).
\]
Thus the corrected increment
\begin{equation}\label{up:eq:corrected-drift}
 Z=\ell+\kappa\{h(\theta')-h(\theta)\}
\end{equation}
satisfies $\E(Z\mid\theta)\ge d\kappa^2/2$ and $|Z|\le C\kappa$
for sufficiently small $\kappa$. Follow the phases of successive
products from any fixed real unit vector $u$. The centered increments
of their corrected log norms are martingale differences bounded by
$C\kappa$. The conditional exponential estimate
\cite[Remark~1]{Azuma} and Chernoff's inequality give
\[
 \Prob\left(
 \log\norm{B_N\cdots B_1u}
 <\frac d4N\kappa^2-2\kappa\norm h_\infty
 \right)\le e^{-cN\kappa^2}.
\]
Since a determinant-one matrix has operator norm at least one, splitting
into this exceptional event and its complement proves
$\E\norm{B_N\cdots B_1}^{-2}\le Ce^{-cN\kappa^2}$.
In the conjugacy $B=S_\kappa(-M)S_\kappa^{-1}$, the condition number
of $S_\kappa$ means $\|S_\kappa\|\|S_\kappa^{-1}\|$, which is at most
$C/\kappa$. Hence
\[
 \|\mathcal P_N\|^{-2}\le C\kappa^{-2}\|B_N\cdots B_1\|^{-2},
\]
proving \eqref{up:eq:near-products}.

For \eqref{up:eq:away-products}, write
$P_E\varphi([u])=\E\varphi([M(E,v_0)u])$ for the projective Markov
operator and
$F(E,[u])=\E\log(\norm{M(E,v_0)u}/\norm u)$.
Both the transition and $F$ are jointly continuous on the compact
energy--projective space. The unipotent argument of
Lemma~\ref{lem:furstenberg} uses only two distinct support points.
The generated group is noncompact and has no finite invariant set of
projective directions; bounded support supplies the logarithmic moment
condition. By \cite[Theorems~8.5--8.6]{Furstenberg}, every stationary measure
$\nu$ at $E\ne0$ has the same strictly positive $F$-average.
The set of pairs $(E,\nu)$ with $E\in K_0$ and $P_E^*\nu=\nu$ is compact.
Hence these stationary averages have a common positive lower bound
$\lambda_*>0$. Stationarity here is with respect to the averaged
operator $P_E$, not each individual matrix.

The expected finite-length log growth is uniformly positive as well.
Otherwise, there would be $N_i\to\infty$, $E_i\to E_*\in K_0$ and starting
directions $u_i$ whose normalized expected growth is smaller than half
that lower bound. A weak limit of
\[
 \nu_i=\frac1{N_i}\sum_{j=0}^{N_i-1}(P_{E_i}^*)^j\delta_{[u_i]}
\]
is stationary at $E_*$: the invariance error against a continuous test
function is at most twice its supremum norm divided by $N_i$.
Moreover, for unit $u_i$,
\[
 N_i^{-1}\E\log\|\mathcal P_{N_i}(E_i)u_i\|
 =\int F(E_i,[u])\,d\nu_i([u]).
\]
Joint continuity then contradicts the lower bound on the $F$-average.
Consequently, there are a fixed integer $m$ and $a_0>0$ such that
\[
 \inf_{E\in K_0,\,\|u\|=1}
 \E\log\|\mathcal P_m(E)u\|\ge a_0.
\]
For a fixed initial unit vector $u$, let $G_j$ be the product of the
cells numbered $jm,\ldots,(j+1)m-1$, in transfer order, and set
$U_j=\mathcal P_{jm}(E)u/\|\mathcal P_{jm}(E)u\|$ and
$X_j=\log\|G_jU_j\|$. The filtration $\mathcal F_{jm}$ contains the
first $jm$ variables. Multiplying the successive norm ratios gives
$\sum_{j=0}^{k-1}X_j=\log\|\mathcal P_{km}(E)u\|$.
Independence of the next block gives
$\E(X_j\mid\mathcal F_{jm})\ge a_0$ for every $U_j$.
There is a uniform $b_0<\infty$ with $|X_j|\le b_0$, since the
one-cell matrices and their inverses are bounded on $K_0\times[a,b]$.
For $k$ complete blocks the centered sum
\[
 S_k=\sum_{j=0}^{k-1}\{X_j-\E(X_j\mid\mathcal F_{jm})\}
\]
has increments bounded by $2b_0$. The conditional exponential bound
\cite[Remark~1]{Azuma} gives
$\E(e^{-t(S_{j+1}-S_j)}\mid\mathcal F_{jm})\le e^{2b_0^2t^2}$.
Iterating and taking $t=a_0/(8b_0^2)$ yields
\[
 \Prob(S_k\le-a_0k/2)
 \le \exp\{-a_0^2k/(32b_0^2)\}.
\]
Writing $N=km+r$, $0\le r<m$, the last $r$ factors can reduce the
log norm by at most $m\log C_0$, where $C_0$ is chosen so that
\[
 \sup_{E\in K_0,\,x\in[a,b]}
       \max\{\|M(E,x)\|,\|M(E,x)^{-1}\|\}\le C_0.
\]
Thus, outside an event of probability $Ce^{-cN}$,
$\log\|\mathcal P_N(E)u\|\ge cN-C$.
On the exceptional event use $\|\mathcal P_N(E)\|^{-2}\le1$;
on its complement use this lower bound. Increasing $C$ for $N<m$
proves \eqref{up:eq:away-products}, uniformly in $E\in K_0$.

Finally, reversal of a finite iid sequence does not change its law, and
$\norm{A^{-1}}=\norm A$ for a $2$ by $2$ matrix with determinant one.
The inverse-product estimates follow. The norm estimate uses a fixed
starting vector; its exceptional event may depend on that vector.
It does not give positive samplewise growth simultaneously in all
directions: for $A\in\SL(2,\R)$,
$\inf_{\|u\|=1}\log\|Au\|=-\log\|A\|\le0$.
\end{proof}

\Needspace*{10\baselineskip}
\section{Upper bounds and exact transport exponents}
\label{sec:upper-transport}
In this section assume \eqref{rt:eq:law} and $\sigma_v^2>0$.
All constants may depend on the law, and are not uniform as its variance
tends to zero. Lemma~\ref{up:lem:products} supplies real-energy growth.
We transfer it to complex energy and obtain the energy-dependent
spatial radius in Proposition~\ref{up:prop:outside}; a weighted Schur
estimate treats the critical window. Extracting the polynomial factors
before Poisson smoothing retains the full threshold index in
Proposition~\ref{up:prop:finite}. The lower bound of
Theorem~\ref{rt:thm:main} then gives Theorem~\ref{up:thm:front}, with
a two-time argument covering the zero-exponent regime.

\subsection{Complex-energy estimates}
\begin{lemma}\label{up:lem:polynomials}
Let $\Phi_I(z)$ be the physical transfer matrix across a finite interval
of a real, hopping-one discrete Schr\"odinger operator, in either
orientation. For real $E$ and $\varepsilon>0$,
\begin{equation}\label{up:eq:polynomial-comparison}
 \norm{\Phi_I(E+i\varepsilon)}
 \ge 2^{-1/2}\norm{\Phi_I(E)}.
\end{equation}
\end{lemma}
\begin{proof}
Write $p_{a,b}(z)=\det(z-H^D_{[a,b]})$, with $p_{a,a-1}=1$ and
$p_{a,a-2}=0$. In the forward orientation, the transfer across $[a,b]$ is
\[
 \Phi_{[a,b]}(z)=
 \begin{pmatrix}
 p_{a,b}(z)&-p_{a+1,b}(z)\\
 p_{a,b-1}(z)&-p_{a+1,b-1}(z)
 \end{pmatrix}.
\]
For a one-site interval, this is
$\left(\begin{smallmatrix}z-V(a)&-1\\1&0\end{smallmatrix}\right)$.
Every nonconstant polynomial in these entries has only real
zeros, because the corresponding finite Jacobi matrix is self-adjoint.
Factoring it over those zeros gives $|p(E+i\varepsilon)|\ge|p(E)|$.
The comparison follows first for the Frobenius norm and then for the
operator norm. Reversing the orientation takes an adjugate and only
permutes the signed entries. This argument concerns the physical
transfer matrix, not its energy-dependent singular conjugate.
\end{proof}

\begin{lemma}\label{up:lem:CT}
Assume the deterministic bounds \eqref{tr:eq:assumption}.
There are $\epsilon_0,c,C>0$, depending only on $a,b$, such that for
$|E|\le\epsilon_0$, $0<\varepsilon\le\epsilon_0$, every $j\in\Z$ and $R\ge1$,
\begin{equation}\label{up:eq:critical-CT}
 \sum_{|n-j|>R}|G(n,j;E+i\varepsilon)|^2
 \le C\varepsilon^{-2}
 \exp\left(-\frac{c\varepsilon R}{\sqrt{|E|+\varepsilon}}\right).
\end{equation}
For $0<E\le\epsilon_0$ one also has
\begin{equation}\label{up:eq:gap-CT}
 \sum_{|n-j|>R}|G(n,j;E+i\varepsilon)|^2
 \le CE^{-2}e^{-c\sqrt E R}.
\end{equation}
\end{lemma}
\begin{proof}
Continue the Schur complement from Section~\ref{sec:gap-schur} to
$z=E+i\varepsilon$:
\[
 c_z=(V-z)^{-1},\qquad Q_z=z+D^*c_zD.
\]
For small $\epsilon_0$, uniformly in the configuration,
$\operatorname{Re}c_z\ge c_*>0$ and
$0\le\operatorname{Im}c_z\le C\varepsilon$.
Take a bounded real function $\phi$ on cells with
$|\phi(m+1)-\phi(m)|\le1$, put $W=e^{t\phi}$ and $Q_t=WQ_zW^{-1}$.
For $d_m=t(\phi(m+1)-\phi(m))$, the form
$\langle u,Q_tu\rangle$ is $z\|u\|^2$ plus the sum over $m$ of
$c_z(m)$ times
\[
 |u_m-u_{m+1}|^2
 -2(\cosh d_m-1)\operatorname{Re}(\overline{u_m}u_{m+1})
 +2i\sinh d_m\operatorname{Im}(\overline{u_m}u_{m+1}).
\]
The sum converges absolutely for $u\in\ell^2$, since the coefficients
and $d_m$ are bounded and $\sum_m|u_mu_{m+1}|\le\|u\|^2$.
For $0\le t\le t_0$, Cauchy's inequality and
$\operatorname{Im}(\overline{u_m}u_{m+1})
=\operatorname{Im}(\overline{u_m}(u_{m+1}-u_m))$ imply
\begin{align}
 \operatorname{Re}\ip{u}{Q_tu}
 &\ge c_1\norm{Du}^2-(|E|+Ct^2)\norm u^2,\label{up:eq:form-real}\\
 \operatorname{Im}\ip{u}{Q_tu}
 &\ge\varepsilon(1-Ct^2)\norm u^2
             -Ct\norm u\,\norm{Du}.\label{up:eq:form-imag}
\end{align}
For $Q_tu=g$, the first inequality yields
\[
 \norm{Du}\le C\sqrt{\norm u\,\norm g}
                  +C\sqrt{|E|+t^2}\,\norm u.
\]
Choose $t=c_0\varepsilon/\sqrt{|E|+\varepsilon}$, with $0<c_0\le1$
depending only on $a,b$. Then
$t^2\le c_0^2\varepsilon$ and
$t\sqrt{|E|+t^2}\le c_0\varepsilon$.
For $u\ne0$, divide \eqref{up:eq:form-imag} by $\|u\|$ and use
$|\ip{u}{g}|\le\|u\|\|g\|$ to obtain
\[
 \varepsilon(1-Ct^2)\|u\|
 \le\|g\|+Ct\sqrt{\|u\|\|g\|}
            +Ct\sqrt{|E|+t^2}\|u\|.
\]
Young's inequality bounds the middle term by
$\varepsilon\|u\|/4+Ct^2\|g\|/\varepsilon$.
First choose $c_0$ so that the last term is at most
$\varepsilon\|u\|/4$, and then choose $\epsilon_0$ so that
$Ct^2\le1/4$ and $t\le t_0$.
Since $t^2/\varepsilon\le c_0^2$, absorption gives
$\|u\|\le C\varepsilon^{-1}\|g\|$, with $C$ independent of
$E,\varepsilon,\phi$ and the configuration.
The Schur complement $Q_z$ is invertible. Since both multiplication
operators $W$ and $W^{-1}$ are bounded, $Q_t=WQ_zW^{-1}$ is invertible
as well; the preceding bound is therefore a resolvent norm estimate.

Schur inversion now gives, in the staggered decomposition,
\[
 (\widetilde H-z)^{-1}=
 \begin{pmatrix}
  -Q_z^{-1}&Q_z^{-1}D^*c_z\\
  c_zDQ_z^{-1}&c_z-c_zDQ_z^{-1}D^*c_z
 \end{pmatrix}.
\]
Conjugating both cell components by $W$ replaces $Q_z^{-1}$ by
$Q_t^{-1}$, $D$ by $D_t=WDW^{-1}$, and $D^*$ by
$D_t^{\sharp}=WD^*W^{-1}$ (not by $D_t^*$).
Both difference operators have norm at most $1+e^t$, and $c_z$ commutes
with $W$ and is uniformly bounded. The four blocks displayed above
therefore each have norm at most $C/\varepsilon$. The norm of the
full two-by-two block operator is at most twice the largest block norm,
so the same bound holds for the full weighted resolvent.

Here cell $m$ is the pair $(2m,2m+1)$. Write
$j=2m_j+\iota_j$ and $n=2m+\iota$, with $\iota,\iota_j\in\{0,1\}$;
then $|n-j|\le2|m-m_j|+1$. Use the bounded weight
$\phi(m)=\min\{|m-m_j|,R\}$. For $R\ge4$, physical sites with
$|n-j|>R$ belong to cells with $\phi(m)\ge(R-2)/2\ge R/4$.
The source weight is one. Writing $\widetilde\delta_j$ for the same unit
source in the staggered coordinates, the weighted column estimate gives
\[
 \sum_{|n-j|>R}|G(n,j;z)|^2
 \le e^{-tR/2}\|\operatorname{diag}(W,W)
                      (\widetilde H-z)^{-1}\widetilde\delta_j\|^2
 \le C\varepsilon^{-2}e^{-tR/2}.
\]
The staggering changes only signs and the ordering of coordinates.
For $1\le R<4$ the unweighted resolvent bound suffices, after increasing
$C$. This proves \eqref{up:eq:critical-CT} without an unbounded-weight
limit.

For $E>0$, retain the sign of the energy in the real form estimate:
\[
 \operatorname{Re}\ip{u}{Q_tu}
 \ge c_1\norm{Du}^2+(E-Ct^2)\norm u^2.
\]
Choose $c_0$ so that $Cc_0^2\le1/2$ and take $t=c_0\sqrt E$.
Then $\|Q_t^{-1}\|\le2/E$ by the real form estimate, and Schur
inversion gives a full conjugated resolvent norm at most $C/E$.
The same truncated weights prove \eqref{up:eq:gap-CT}.
\end{proof}

\subsection{Outside probabilities at fixed energy}
For a finitely supported state $f$, set
\[
 P_f(R,T)=\frac2T\int_0^\infty e^{-2t/T}
       \sum_{|n|>R}|(e^{-itH_\omega}f)(n)|^2\,dt.
\]
Choose $K\ge4$ with $\spec H_\omega\subset[-K+1,K-1]$ for every
configuration under consideration. The following form of the
full-line estimate keeps the physical-site and cell conventions explicit.
\begin{lemma}\label{up:lem:DT}
For $|E|\le K$, $0<\varepsilon\le1$, $j\in\Z$ and $L\ge2$,
\begin{equation}\label{up:eq:DT-interface}
 \sum_{|n-j|>4L+2}|G(n,j;E+i\varepsilon)|^2
 \le C_K\varepsilon^{-4}
       \sum_{\sigma\in\{+,-\}}\norm{\Phi_{j-1}(\sigma2L,E+i\varepsilon)}^{-2}.
\end{equation}
Here $\Phi_k(n,z)$ maps $(u(k+1),u(k))^t$ to
$(u(k+n+1),u(k+n))^t$. The constant is uniform in the source and the
configuration.
\end{lemma}
\begin{proof}
For source $1$, \cite[Lemma~3]{DTUpper} bounds the right Green
tail beyond $N\ge3$ by
$C_K\varepsilon^{-4}(\max_{3\le n\le N}\|\Phi_0(n,z)\|^2)^{-1}$.
Take $N=4L+2$ and retain $n=2L$ in the maximum; translation by $j-1$
gives the right-tail bound with base $j-1$. Restricting to the slightly
smaller tail $n-j>4L+2$ preserves the estimate.

For the left tail reflect about the source, so the reflected coordinate
is $m=j+1-n$. If $P=\matr{0}{1}{1}{0}$, the reflected forward transfer
of length $2L$ is $P\Phi_j(-2L,z)P$. Its base differs from the one in
\eqref{up:eq:DT-interface} by one site:
\[
 \Phi_j(-2L,z)
 =T(z,V(j-2L))\Phi_{j-1}(-2L,z)T(z,V(j))^{-1}.
\]
These end factors and their inverses are uniformly bounded for
$|E|\le K$, $\varepsilon\le1$. Reflection therefore gives the left
bound with the same type of constant, as in the two-sided application
in \cite[Theorem~7]{DTUpper}. Adding the two bounds proves the lemma.

For later use, put $T_0=T(z,0)$ and
$\mathcal P_{r,L}(z)=M(z,v_{r+L-1})\cdots M(z,v_r)$.
The parity conversion is exactly
\[
 \Phi_{2r}(2L,z)=\mathcal P_{r,L}(z),\qquad
 \Phi_{2r-1}(2L,z)=T_0^{-1}\mathcal P_{r,L}(z)T_0.
\]
Backward transfers are the inverses of the corresponding shifted
forward products. Thus both parities reduce to $L$ cell factors with
uniformly bounded end conjugacies. The source and the Green function
remain full-line objects; no half-line spectral measure is used.
\end{proof}

\begin{lemma}\label{up:lem:mass}
Assume \eqref{tr:eq:assumption}, and let
$\mu_j(I)=\ip{\delta_j}{\mathbf1_I(H)\delta_j}$.
Put $a_j=1/2$ for even $j$ and $a_j=3/2$ for odd $j$.
Uniformly in the configuration and the source, for small $\delta>0$,
\begin{equation}\label{up:eq:mass}
 \mu_j([-\delta,\delta])\le C\delta^{a_j}.
\end{equation}
If $0<\varepsilon\le\delta$, then
\begin{equation}\label{up:eq:poisson-mass}
 \frac{\varepsilon}{\pi}\int_{-2\delta}^{2\delta}
       \norm{R(E+i\varepsilon)\delta_j}^2\,dE
 \le C\delta^{a_j}.
\end{equation}
\end{lemma}
\begin{proof}
The deterministic gap $(0,a)$ and the spectral theorem give
\[
 \mu_j([-\delta,\delta])
 \le4\delta^2\norm{R(\delta)\delta_j}^2.
\]
Lemma~\ref{tr:lem:columns} proves \eqref{up:eq:mass}.
By increasing $C$, the mass estimate holds for every radius.
For \eqref{up:eq:poisson-mass}, integrate the Poisson kernel against
$\mu_j$. Spectral energies $|\lambda|\le4\delta$ contribute at most
$C\delta^{a_j}$. For $|\lambda|>4\delta$ the integrated kernel is bounded
by $C\varepsilon\delta/\lambda^2$. Splitting into dyadic annuli gives
\[
 \varepsilon\delta\int_{|\lambda|>4\delta}\lambda^{-2}\,d\mu_j(\lambda)
 \le C\varepsilon\delta^{a_j-1}
       \sum_{k\ge0}2^{k(a_j-2)}
 \le C\delta^{a_j}.
\]
Here $a_j<2$ and $\varepsilon\le\delta$ are essential. The argument
retains the spectral mass, as in the energy decomposition used for
elliptic polymer models in \cite[Section~6]{JSUpper}; the two exponents
here follow from the mosaic gap column bounds.
\end{proof}

For $0<\rho<1/3$, define the energy-dependent radius
\[
 r_T(E)=
 \begin{cases}
 T^\rho/(|E|+T^{-2/3}),&|E|\le\epsilon_0,\\
 T^\rho,&|E|>\epsilon_0.
 \end{cases}
\]
Its smallest value is at least $cT^\rho$ and its largest is
$T^{2/3+\rho}$.
\begin{proposition}\label{up:prop:outside}
For a sufficiently large fixed $C_0$ and each fixed $j$, set
\[
 \mathcal W_j(T,\rho)=\frac1{\pi T}\int_\R
   \sum_{|n-j|>C_0r_T(E)}|G(n,j;E+i/T)|^2\,dE.
\]
There are $c,C>0$ such that
\begin{equation}\label{up:eq:outside-mean}
 \E\mathcal W_j(T,\rho)\le CT^5e^{-cT^\rho}.
\end{equation}
On one event of probability one, for every $j\in\Z$, every rational
$\rho\in(0,1/3)$, and every $A>0$,
\begin{equation}\label{up:eq:dyadic-remainder}
 \mathcal W_j(2^k,\rho)=O(2^{-kA}).
\end{equation}
\end{proposition}
\begin{proof}
Set $\varepsilon=T^{-1}$ and $\Delta=T^{-2/3}$.
If $|E|\le\Delta$, then $r_T(E)\ge T^{2/3+\rho}/2$, so
\eqref{up:eq:critical-CT} bounds the Green tail by
$CT^2e^{-cT^\rho}$.
If $\Delta<E\le\epsilon_0$, then
$\sqrt E\,r_T(E)\ge cT^\rho$, and \eqref{up:eq:gap-CT} gives a bound
$CT^{4/3}e^{-cT^\rho}$.
If $-\epsilon_0\le E<-\Delta$, use
\eqref{up:eq:DT-interface} with $L=\lfloor r_T(E)\rfloor$.
For large $T$, uniformly in $E$, $r_T(E)\ge4$, so
$L\ge r_T(E)/2$ and $4L+2\le5r_T(E)$.
Since $r_T(E)|E|\ge T^\rho/2$,
Lemmas~\ref{up:lem:products} and~\ref{up:lem:polynomials} give
\[
 \E\norm{\Phi_{j-1}(\sigma 2L,E+i/T)}^{-2}
 \le C|E|^{-1}e^{-cr_T(E)|E|}
 \le CT^{2/3}e^{-cT^\rho}.
\]
Taking $C_0$ large absorbs the integer rounding and the fixed enlargement
in the full-line estimate.
For $\epsilon_0\le|E|\le K$, use \eqref{up:eq:away-products} at distance
of order $T^\rho$. To handle the unbounded energy range explicitly,
put $B_0=2+b$ and enlarge $K$ so that $K\ge B_0+1$.
For $|E|>K$, let $d_E=|E|-B_0\ge1$. With the physical-site weight
$W_R(n)=\exp(\alpha\min\{|n-j|,R\})$, choose $\alpha>0$ such that
$2(e^\alpha-1)\le1/2$. Then
$\|W_RHW_R^{-1}-H\|\le1/2\le d_E/2$ and the resolvent identity gives
$\|W_RR(E+i/T)W_R^{-1}\|\le2/d_E$.
Consequently the tail beyond $C_0T^\rho$ is at most
$4d_E^{-2}e^{-2\alpha C_0T^\rho}$. Its integral with factor
$1/(\pi T)$ is bounded by
\[
 \frac{8}{\pi T(K-B_0)}e^{-2\alpha C_0T^\rho}.
\]
Integration with the factor $1/(\pi T)$ proves
\eqref{up:eq:outside-mean}; its polynomial prefactor is not meant to
be sharp.

For a positive integer $B$, Markov's inequality gives
$\Prob(\mathcal W_j(2^k,\rho)>2^{-kB})
\le C2^{k(B+5)}e^{-c2^{k\rho}}$, which is summable in $k$.
Borel--Cantelli gives a finite starting index depending on the sample,
$j,\rho,B$. Intersect over integer $j$,
rational $\rho\in(0,1/3)$, and positive integer powers. Increasing
the power gives \eqref{up:eq:dyadic-remainder} for all $A>0$.
Only dyadic times are needed here; the moment estimate below is extended
to all times after energy integration.
\end{proof}

\subsection{Moment exponents}
We apply the polynomial tail identity before Poisson smoothing.
Lemma~\ref{up:lem:mass} smooths the two basic point-source measures,
whose threshold powers are below two. Factoring out $A_f(z)$ and
$B_f(z)$ first retains their vanishing orders even when
$\nu_H(f)/2>2$. The next proposition carries out this step and closes
the upper half of Theorem~\ref{up:thm:front}.
\begin{proposition}\label{up:prop:finite}
On one event of probability one, for every nonzero finitely supported $f$, every
$q>0$, and every $\rho>0$, there are sample-dependent constants such that
\begin{equation}\label{up:eq:finite-moments}
 \M_q(T;f)\le C_{f,q,\rho}T^{\rho q}(1+\log T)
       \bigl(1+T^{(2q-\nu_H(f))/3}\bigr)
 \qquad(T\ge T_0).
\end{equation}
\end{proposition}
\begin{proof}
Work on the event of Proposition~\ref{up:prop:outside}, first with
$T=2^k$ and rational $0<\rho<1/3$.
Let $S_f=\supp f$. For large $T$, the fixed spatial offsets in $S_f$
are absorbed by doubling $C_0$, since $r_T(E)\ge cT^\rho$ uniformly
in $E$. Cauchy's inequality gives the unweighted estimate
\[
\begin{aligned}
 U_f(T)&:=\frac1{\pi T}\int_\R
       \sum_{|n|>2C_0r_T(E)}|(R(E+i/T)f)(n)|^2\,dE\\
 &\le\|f\|^2\sum_{j\in S_f}\mathcal W_j(T,\rho)
 =O_{f,B}(T^{-B})\qquad(B>0).
\end{aligned}
\]
The event is unchanged; the threshold time may depend on the finite set
$S_f$. This estimate controls the middle part of the weighted tail below.

The hopping bound in Lemma~\ref{tr:lem:parseval}, together with unitarity,
gives the deterministic estimate
\begin{equation}\label{up:eq:ballistic-tail}
 P_f(R,T)\le C_f e^{-cR/T}\qquad(T\ge1,\ R\ge R_f),
\end{equation}
Indeed, fix $\alpha>0$. Summing the squared hopping bound outside
$|n|>R$ and absorbing the finite support of $f$ gives
\[
 \sum_{|n|>R}|(e^{-itH}f)(n)|^2
 \le C_f\exp(4e^\alpha t-2\alpha R).
\]
Choose $0<\eta\le\alpha/(4e^\alpha)$ and split the time integral
at $t=\eta R$. The earlier part is at most $C_fe^{-\alpha R}$,
since the averaging measure has total mass one. Unitarity bounds
the later part by $\|f\|^2e^{-2\eta R/T}$. For $T\ge1$ their sum
implies \eqref{up:eq:ballistic-tail}.
To pass from mass to moments, put
\[
 a_T(n)=\frac2T\int_0^\infty e^{-2t/T}|(e^{-itH}f)(n)|^2\,dt.
\]
Tonelli's theorem and $x^q=R^q+q\int_R^x s^{q-1}\,ds$ for $x>R$
give the exact identity
\[
 \begin{aligned}
 \sum_{|n|>R}|n|^q a_T(n)
 &=R^qP_f(R,T)+q\int_R^\infty s^{q-1}P_f(s,T)\,ds\\
 &\le C_{f,q}(R^q+T^q)e^{-cR/(2T)}.
 \end{aligned}
\]
At $R=T^2$ this is at most $C_{f,q}T^{2q}e^{-cT/2}$.
By Parseval, it also bounds the resolvent contribution from $|n|>T^2$
after integration over all real $E$. On the remaining part
$2C_0r_T(E)<|n|\le T^2$, the weight is at most $T^{2q}$ and the
integrated mass is at most $U_f(T)$. Choosing $B>A+2q$ above proves
\[
 \frac1{\pi T}\int_\R\sum_{|n|>2C_0r_T(E)}
       |n|^q|(R(E+i/T)f)(n)|^2\,dE=O_{f,q,A}(T^{-A})
 \qquad(A>0).
\]

It remains to estimate the inner contribution.
The fixed exceptional set $F_f$ from \eqref{tr:eq:tail-module} contributes
at most $\max_{n\in F_f}|n|^q\norm f^2$ to the entire time-averaged
moment. Outside that set, write $r=r_f$, $s=s_f$, and
$\nu=\nu_H(f)$. On $|E|\le2\delta$, with $T^{-1}\le\delta\le\epsilon_0$,
\[
 |A_f(E+i/T)|\le C_f\delta^r,\qquad
 |B_f(E+i/T)|\le C_f\delta^s,
\]
with zero polynomials omitted. The tail identity and
Lemma~\ref{up:lem:mass} therefore give
\begin{equation}\label{up:eq:channel-mass}
 \frac1{\pi T}\int_{-2\delta}^{2\delta}
 \sum_{n\notin F_f}|(R(E+i/T)f)(n)|^2\,dE
 \le C_f\bigl(\delta^{2r+1/2}+\delta^{2s+3/2}\bigr)
 \le C_f\delta^{\nu/2}.
\end{equation}
This use of the two point-source measures is valid even when $\nu/2>2$.

Cover $|E|\le\epsilon_0$ by the central window $|E|\le2\Delta$,
$\Delta=T^{-2/3}$, and dyadic annuli
$\delta\le|E|\le2\delta$, where $\delta=2^\ell\Delta\ge\Delta$.
The central radius is at most $CT^\rho/\Delta$; on each annulus the
inner radius is at most $CT^\rho/\delta$.
By \eqref{up:eq:channel-mass}, their total contribution is bounded by
\[
 C_fT^{\rho q}\sum_{\substack{\delta=2^\ell\Delta\\
                         \Delta\le\delta\le\epsilon_0}}
                 \delta^{\nu/2-q}
 \le C_{f,q}T^{\rho q}(1+\log T)
                  \bigl(1+T^{(2q-\nu)/3}\bigr).
\]
A bounded overlap at the first or last window changes only the constant.
More precisely, for $s=\nu/2-q$ the geometric sum
$\sum_{\ell=0}^{\lfloor\log_2(\epsilon_0/\Delta)\rfloor}(2^\ell\Delta)^s$
is $O(1)$ if $s>0$, $O(1+\log T)$ if $s=0$, and
$O(\Delta^s)=O(T^{(2q-\nu)/3})$ if $s<0$.
Thus only $2q=\nu$ requires the logarithm; the displayed proposition
uses one bound for all three cases.
For $|E|>\epsilon_0$, the inner radius is $CT^\rho$; unweighted
Parseval bounds that contribution by $CT^{\rho q}\norm f^2$.
This proves the estimate on dyadic times.

For $2^k\le T\le2^{k+1}$, positivity gives
$\M_q(T;f)\le(2^{k+1}/T)\M_q(2^{k+1};f)$.
The prefactor is at most two, so the same powers hold for every large
$T$. Given an arbitrary real $\rho>0$, use a smaller positive rational
value. The event is independent of $f$ and $q$; only the constants and
the starting time depend on them.
\end{proof}

\begin{proof}[Proof of Theorem~\ref{up:thm:front}]
Proposition~\ref{up:prop:finite}, with $\rho\downarrow0$, gives
\[
 \beta_f^+(q)\le\max\left\{0,\frac23-\frac{\nu_H(f)}{3q}\right\}.
\]
Intersect its event with the event of Theorem~\ref{rt:thm:main}, which
gives the lower bound $2/3-\nu_H(f)/(3q)$.

For completeness, the lower exponent is also nonnegative, including at
$2q=\nu_H(f)$. Choose $s>0$ with
$c_s=|\ip{\delta_0}{e^{-isH}\delta_0}|<1$; such a choice exists because
$H\delta_0$ is not a multiple of $\delta_0$.
Indeed a Taylor expansion gives
$|\langle\delta_0,e^{-isH}\delta_0\rangle|^2=1-2s^2+O(s^3)$,
since $H\delta_0=\delta_{-1}+\delta_1$.
For two unit vectors $x,y$, the map
$L:\C^2\to\ell^2$, $L(a,b)=ax+by$, has Gram matrix with eigenvalues
$1\pm|\langle x,y\rangle|$. Hence
$|\langle x,g\rangle|^2+|\langle y,g\rangle|^2
\le(1+|\langle x,y\rangle|)\|g\|^2$.
Apply this with $x=\delta_0$, $y=e^{isH}\delta_0$ and
$g=e^{-itH}f$ to obtain
\[
 |\ip{\delta_0}{e^{-itH}f}|^2+
 |\ip{\delta_0}{e^{-i(t+s)H}f}|^2
 \le(1+c_s)\norm f^2.
\]
Put $b_f(t)=\sum_{n\ne0}|\ip{\delta_n}{e^{-itH}f}|^2$.
Then $0\le b_f(t)\le\|f\|^2$ and the preceding inequality gives
$b_f(t)+b_f(t+s)\ge(1-c_s)\|f\|^2$.
Define
\[
 B_T=\frac2T\int_0^\infty e^{-2t/T}b_f(t)\,dt.
\]
The exact shift identity gives
\[
 \begin{aligned}
 \frac2T\int_0^\infty e^{-2t/T}b_f(t+s)\,dt
 &=e^{2s/T}\left(B_T-\frac2T\int_0^s e^{-2u/T}b_f(u)\,du\right)\\
 &=B_T+O_{s,f}(T^{-1}).
 \end{aligned}
\]
Here boundedness of $b_f$, rather than a bound on the growing moment,
justifies the error. Averaging the two-time inequality and using
$\M_q(T;f)\ge B_T$ now yields
\[
 \M_q(T;f)\ge
 \frac{1-c_s}{2}\norm f^2-O_{s,f}(T^{-1}),
\]
since $|n|^q\ge1$ for $n\ne0$.
Thus $\beta_f^-(q)\ge0$. The upper and lower bounds agree for every
$q>0$. The common event works for all finite states, and sending
$q\to\infty$ gives $\alpha_u^\pm(f)=2/3$.
\end{proof}

\begin{remark}\label{up:rem:periodic}
If the law is a point mass at $v>0$, the two-band Bloch matrix is
\[
 H(\theta)=
 \begin{pmatrix}0&1+e^{-i\theta}\\1+e^{i\theta}&v\end{pmatrix},
 \qquad
 E_\pm(\theta)=\frac v2\pm
       \sqrt{\frac{v^2}{4}+4\cos^2(\theta/2)}.
\]
The bands are separated. For a nonzero finitely supported state, its Bloch transform
is a nonzero trigonometric-polynomial vector. The fiber evolution is
$\sum_\sigma e^{-itE_\sigma}P_\sigma\widehat f$. Applying $i\partial_\theta$
gives the leading term
$t\sum_\sigma e^{-itE_\sigma}E_\sigma'P_\sigma\widehat f$ and a remainder
bounded in $L^2$ uniformly in $t$. Orthogonality of the two projections
shows that the leading squared-norm coefficient is
\[
 \int_0^{2\pi}\sum_{\sigma\in\{+,-\}}
 |E_\sigma'(\theta)|^2
 \norm{P_\sigma(\theta)\widehat f(\theta)}^2\,\frac{d\theta}{2\pi}>0.
\]
The positivity follows because the derivatives vanish only at finitely
many angles. Since physical position is twice cell position plus
a bounded sublattice term, the physical second moment is four times
this coefficient times $t^2$, with error $O(t+1)$.
Finally $(2/T)\int_0^\infty e^{-2t/T}t^2\,dt=T^2/2$, so
$\M_2(T;f)=c_fT^2+O(T)$ with $c_f>0$.
Thus the nondegeneracy assumption in Theorem~\ref{up:thm:front} cannot
be dropped. This does not contradict the lower bounds, which include
constant laws.
\end{remark}

\subsection{Further questions}
\label{sec:conclusion}
Theorem~\ref{up:thm:front} separates the propagation scale from the
initial state's coupling to the threshold. The former gives the
large-moment exponent $2/3$; the latter determines the correction at
each finite moment order. The assumptions include strictly positive
nondegenerate Bernoulli laws, while the deterministic lower bound
requires no randomness. None of these statements classifies spectral
type.

The remaining questions concern information not determined by the
exponents: leading moment constants, subpower growth when
$2q\le\nu_H(f)$, and pointwise-in-time behavior. It is also open in this
argument whether the double logarithm in the almost-sure lower bound
can be removed. For larger mosaic periods, a transport formula would
have to account for the initial state's coupling to several critical
energies; the present two-channel classification does not assert such
a formula.
\section*{Acknowledgements}
One of the authors (F.~N.) would like to thank Jake Fillman and
Hermann Schulz-Baldes for their useful discussions and valuable comments.

\section*{Statements and declarations}
\subsection*{Funding}
Fumihiko Nakano acknowledges support from JSPS KAKENHI Grant Number JP26K00608.
Xu Xia was partially supported by the National Natural Science Foundation
of China (Grant No.~12301218).
\subsection*{Data availability}
This article is theoretical and no datasets were generated or analysed during
the current study. Auxiliary finite checks used during revision are not inputs
to any proof and are available from the corresponding author upon reasonable
request.
\subsection*{Competing interests}
The authors declare that they have no competing interests.

\begingroup
\setlength{\emergencystretch}{1em}
\urlstyle{same}

\endgroup
\end{document}